\documentclass[a4paper,11pt]{article}
\usepackage{jcappub} % for details on the use of the package, please
\usepackage[T1]{fontenc} % if needed

\usepackage{dcolumn}
\usepackage{bm}
\usepackage{graphics}
\usepackage{graphicx}
\graphicspath{{figures/}}
\usepackage{epsfig}
\usepackage{booktabs,multirow}
\usepackage{hhline}
\usepackage{slashed}
\usepackage{rccol}
\usepackage{mathrsfs}
\usepackage{amsmath, amssymb}
\usepackage[dvipsnames]{xcolor,colortbl}
\usepackage{subfigure}
\newcommand{\orcid}[1]{\href{https://orcid.org/#1}{\includegraphics[width=10pt]{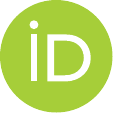}}}

\newcommand*{\SavedEqref}{}
\let\SavedEqref\eqref
\renewcommand*{\eqref}[1]{%
	\begingroup
	\hypersetup{
		linkcolor=linkequation,
		linkbordercolor=linkequation,
	}%
	\SavedEqref{#1}%
	\endgroup
}

\title{Signatures of $X_{17}$ through Coherent Elastic Solar Neutrino-Nucleus Scattering in Direct Detection Searches}

\author[a]{M. F.~Mustamin\,\orcid{0000-0003-3996-4651},}
\author[a,1]{M.~Demirci\,\orcid{0000-0003-2504-6251}%
\note{Corresponding Author},}
\author[b]{and M.~Deniz\,\orcid{0000-0002-9781-8241}}
\affiliation[a]{Department of Physics, Karadeniz Technical University, Trabzon 61080, Türkiye}
\affiliation[b]{Department of Physics, Dokuz Eylül University, Buca,
	Izmir 35160, Türkiye}

\emailAdd{mfmustamin@ktu.edu.tr}
\emailAdd{mehmetdemirci@ktu.edu.tr}
\emailAdd{muhammed.deniz@deu.edu.tr}
\date{\today}
\abstract{
The $X_{17}$ particle has been proposed to explain the invariant mass anomalies observed in electron-positron pairs during nuclear transitions at the Atomki experiment. Motivated by recent observations of $^8$B solar neutrinos induced coherent elastic neutrino-nucleus scattering (CE$\nu$NS), we present the first comprehensive analysis of the hypothetical boson using data from multi-ton dark matter direct detection facilities.
We consider the new particle as a light $Z'$ mediator arising from a spontaneously broken $U(1)'$ symmetry, featuring both vector and axial-vector couplings to leptons.
By evaluating the latest datasets from XENONnT, PandaX-4T, and LUX-ZEPLIN, we derive stringent limits on the effective vector coupling utilizing marginalization procedures.
Our global analysis provides competitive constraints that meaningfully narrow the allowed parameter space of the model, while exhibiting a clear sensitivity to the tau-neutrino coupling.}


\begin{document}
\maketitle
\flushbottom
	
\section{Introduction}
Recent anomalies at the Atomki laboratory \citep{Gulyas:2015mia} have received considerable experimental and theoretical attention.
An unexpected excess of electron-positron pair creation emerged from the decay of an excited $^8$Be at large opening angles, with a confidence level more than $5\sigma$, was observed in nuclear transition processes with an invariant mass of approximately $16.7$ MeV~\citep{Krasznahorkay:2015iga}.
The finding was repeated using $^4$He \citep{Krasznahorkay:2021joi} and $^{12}$C \citep{Krasznahorkay:2022pxs} with similar results. Many other experiments attempted to verify this finding with reported consistent results~\citep{Anh:2024req,Abraamyan:2024bbs,PADME:2025dla}, while others disfavoured the anomaly~\citep{MEGII:2024urz}. 
Furthermore, a satisfactory explanation cannot be provided by standard nuclear physics or strong force approaches \citep{Zhang:2017zap,Koch:2020ouk,Hayes:2021hin,Viviani:2024czq}.
The Atomki collaboration conjectures that the anomalous signal does not originate from Standard Model (SM) effects, but from a new light boson beyond the SM (BSM).
A new boson termed X17 was accordingly introduced as an explanation of this anomaly. 

The validity of the anomalies and the nature of the state are yet to be fully understood. 
Theoretical interpretations of $X_{17}$ vary, with proposals including a protophobic vector boson or a mediator between visible and dark matter. 
After careful analyses of a variety of scenarios, the data appear to be either a vector mediator \citep{Feng:2016jff,Feng:2016ysn,Feng:2020mbt} or an axial-vector mediator \citep{Kozaczuk:2016nma, Barducci:2022lqd, Hostert:2023tkg, Mommers:2024qzy}. A combination of mixed vector and axial-vector is also recently studied \citep{Fieg:2026zkg, Jiang:2026snq}.
The difference depends on the exact treatment of other datasets and our understanding of nuclear physics.
It is worth mentioning that the scalar type of interaction had been excluded by the $^8$Be and $^4$He, while the pseudoscalar type by the $^{12}$C Atomki results due to parity conservation.  
If confirmed, such a discovery would provide a new force carrier in the MeV regime and a possible connection to the dark sector. 
In this work, we interpret the Atomki anomaly with a light $Z'$ mediator from a $U(1)'$ extension of the SM \citep{DelleRose:2017xil,DelleRose:2018eic,Pulice:2019xel} that features vector and axial-vector couplings to all leptons. 
As anticipated in earlier studies \citep{Cerdeno:2016sfi,Boehm:2020ltd,Demirci:2026nju}, such a mediator is expected to play a central role in testing neutrino interactions and BSM physics.

Neutrino scattering processes offer a potential testing ground for the $X_{17}$ boson. Notably, the hypothetical particle signature can potentially be tested in the coherent elastic neutrino nucleus scattering (CE$\nu$NS) \citep{Freedman:1973yd,Drukier:1984vhf}.
After its successful observations by COHERENT \citep{COHERENT:2017ipa, COHERENT:2020iec, COHERENT:2024axu} and later measurements from reactors by Dresden-II \citep{Colaresi:2022obx} and CONUS+\citep{Ackermann:2025obx}, the CE$\nu$NS framework provides strong bounds on new light mediators, which couple to neutrinos and neutrons \citep{Denton:2018xmq, Khan:2019cvi, AristizabalSierra:2022axl, AtzoriCorona:2022moj, Demirci:2023tui, DeRomeri:2025nkx}.
Particularly, several studies have worked out the possibility of observing $X_{17}$ using the CE$\nu$NS framework. In Ref.~\citep{Denton:2023gat}, relevant constraints for the particle were utilized using CE$\nu$NS from reactor antineutrinos. Meanwhile, Ref.~\citep{Cederkall:2025bka} studied the potential observability of $X_{17}$ in the future European Spallation Source \citep{Garoby:2017vew,Baxter:2019mcx,Abele:2022iml}.

In this work, we are particularly interested in exploiting recent results from direct detection searches regarding CE$\nu$NS signals induced by $^8$B solar neutrinos to study the $X_{17}$ particle.
Even though the main purpose of such facilities is to observe dark matter particles, similar observational signatures of neutrino-nucleus scattering make it challenging to eliminate the process completely in practice. 
Using time projection chamber technology, the CE$\nu$NS signal induced by solar neutrinos can be observed, especially from $^8$B and a subdominant hep neutrinos. 
Recent measurements from XENONnT \citep{XENON:2024ijk}, PandaX-4T \citep{PandaX:2024muv}, and LUX-ZEPLIN \citep{LZ:2025igz} reported signals of the CE$\nu$NS induced by $^8$B solar neutrinos,
where the background-only hypothesis is disfavored with a statistical significance of $2.73\sigma$, $2.64\sigma$, and $4.5\sigma$, respectively.
These represent the first observation of solar neutrino-induced nuclear recoils in direct detection searches. 
%as well as the first CE$\nu$NS evidence on a xenon target.
Note that these achievements give rise to a neutrino fog in those experiments, which presents important challenges to discriminate between dark matter and neutrino signals. This fact offers a novel environment of using CE$\nu$NS signals at these facilities to test BSM scenarios, including the $X_{17}$ particle, our main objective in this work. The robust data are expected to provide novel insight into the new boson.

The remainder of this paper is organized as follows: Sec.~\ref{sec:theo} gives a brief explanation of the CE$\nu$NS process and the $Z'$ model responsible for $X_{17}$. Details of the data analysis are presented in Sec.~\ref{section:data}. After that, the predicted event rates are illustrated, and the analysis results are presented in Sec.~\ref{sec:results}. Finally, we conclude in Sec.~\ref{sec:conc}.

\section{Theoretical Framework}\label{sec:theo}
\subsection{CE$\nu$NS in the SM}
CE$\nu$NS constitutes a crucial neutrino-matter interaction channel within the SM, operating as a neutral-current event mediated by the Z boson. During this interaction, a low-energy neutrino scatters off an intact nucleus coherently, imparting a small but distinct kinetic recoil energy. While the experimental detection of such faint nuclear recoils is inherently difficult, the coherent response of the nucleus provides an enhancement to the interaction rate. Specifically, the scattering cross-section demonstrates an approximate dependence on the square of the neutron number, making it the dominant neutrino interaction mechanism at low energy regimes.
The differential cross-section of this process, expressed as a function of the nuclear recoil energy $T_\text{nr}$, can be written as
\begin{align} 
\begin{split}
    \left[\frac{d\sigma_{\nu_\ell}}{dT_\text{nr}}\right]_\text{SM} = \frac{G_F^2 m_\mathcal{N}}{\pi} \left(Q^{\ell,V}_\text{SM}\right)^2 \left(1 - \frac{m_\mathcal{N} T_\text{nr}}{2E_\nu^2} - \frac{T_\text{nr}}{E_\nu}\right) %|F(|\vec{q}|^2)|^2
\end{split}
\label{eq:sm_cevns}
\end{align}
where $G_F$ denotes the Fermi constant, $m_\mathcal{N}$ is nucleus mass, $E_\nu$ is initial neutrino energy, and $\ell=e,\mu,\tau$ represents the neutrino flavor. The SM weak charge is defined by
\begin{align}
Q_\text{SM}^{\ell,V} =g_{\ell,V}^p Z F_Z(|\vec{q}|^2)  + g_V^n N F_N(|\vec{q}|^2) ,
\end{align}
where $Z$ and $N$ represent the proton and neutron numbers of the nucleus, respectively. The vector couplings $g_{\ell,V}^p$ and $g_V^n$ parameterize the weak neutral-current interactions of protons and neutrons, respectively. At leading order, these couplings are given by
\begin{align}
    g_{\ell,V}^p=\left(\frac{1}{2}-2\sin^2\theta_W\right) \quad \text{and} \quad g_V^n=-\frac{1}{2}.
\end{align}
More precise values can be determined by including radiative corrections~\citep{Erler:2013xha, AtzoriCorona:2025xwr}.
We adopt the numerical values of the couplings that correspond to
\begin{align}
g_{e,V}^p = 0.0379, g_{\mu, V}^p = 0.0297, g_{\tau,V}^p= 0.0253
\end{align}
for the proton, and 
\begin{align}
g_V^n = -0.5117
\end{align}
for the neutron. It is clearly seen that the contribution of radiative corrections depends on the neutrino flavor, that introduce a flavor dependence in the neutrino-proton coupling alone.

Moreover, the form factor is denoted by $F(|\vec{q}|^2)$, and we have considered the proton and neutron form factors to be the same. We consider the case when both form factors are given by the Klein-Nystrand parametrization \citep{Klein:1999qj}
\begin{align}
F_Z(|\vec{q}|^2) = F_N(|\vec{q}|^2) = \frac{3j_1(|\vec{q}| R_A)}{|\vec{q}|R_A} \left(\frac{1}{1+|\vec{q}| a_k^2}\right),
\end{align}
where $j_1(x)=\sin(x)/x^2-\cos(x)/x$ is the first order of the spherical Bessel function, $a_k=0.7$ fm, and $R_A=1.23 A^{1/3}$ fm is the root-mean-square radius, with $A$ denoting the atomic mass number. Since the typical momentum transfer of $^8$B solar neutrinos induced CE$\nu$NS process is $|\vec{q}|\sim \mathcal{O}(10)$ MeV, the nuclear form factor dependence is expected to be small.

Recognizing that xenon naturally occurs as a multi-isotope mixture, we evaluate the cross section using the isotopic distribution $(Z,N)_\text{Xe}=[54,(70,72,74,75,$ $76,77,78,80.82)]$, appropriately weighted by their respective natural abundances \citep{Berglund:2009}
\begin{align}
\begin{split}
&f(^{124}\text{Xe})=0.000952,\quad f(^{126}\text{Xe})=0.000890, \\
&f(^{128}\text{Xe})=0.019102,\quad f(^{129}\text{Xe})=0.264006, \\
&f(^{130}\text{Xe})=0.040710,\quad f(^{131}\text{Xe})=0.212324, \\
&f(^{132}\text{Xe})=0.269086,\quad f(^{134}\text{Xe})=0.104357, \\
&f(^{136}\text{Xe})=0.088573.
\end{split}
\end{align}

Neutrinos coming from the Sun can trigger neutrino-nucleus scattering events in dark matter direct detection experiments as the incoming neutrino interacts with the detector target. CE$\nu$NS, together with neutrino-electron scattering, is among the neutrino backgrounds in such experiments that can be used to study BSM physics. In particular, the CE$\nu$NS is sensitive to the $^8$B solar neutrinos, which can significantly boost the detection rate that is proportional to the square of the atomic number of the detector target. This process has a strong degeneracy with the corresponding dark matter-nucleus scattering, which has a very similar nuclear recoil signal. Understanding the CE$\nu$NS channel would help us better separate it from the dark matter signature. Moreover, any deviation from the SM expectation could indicate the existence of new physics at low energy, while the lack of deviation could help to set constraints on the $X_{17}$ particle. 

\subsection{The $X_{17}$ boson in a $U(1)'$ extension}
%\subsection{A $U(1)'$ extension for the $Z'$ as the $X_{17}$}
Extending the SM with a $U(1)'$ gauge group leads to the following Lagrangian \citep{DelleRose:2018eic}
\begin{align}
\mathcal{L} \supset -\frac{1}{4} \hat{F}_{\mu\nu}\hat{F}^{\mu\nu} - \frac{1}{4}\hat{X}_{\mu\nu}\hat{X}^{\mu\nu} - \frac{\kappa}{2} \hat{F}_{\mu\nu}\hat{X}^{\mu\nu},
\end{align} 
where $\hat{F}_{\mu\nu}$ is the SM field tensor and $\hat{X}_{\mu\nu}$ the $U(1)'$ field tensor. The parameter $\kappa$ denotes kinetic mixing between the hypercharge gauge boson $\hat{B}_\mu$ of the SM and the new gauge boson $\hat{X}_\mu$ of the $U(1)'$. Diagonalization can be proceeded by using
\begin{align}
\begin{pmatrix}
\hat{B}_\mu \\ \hat{X}_\mu 
\end{pmatrix} = \begin{pmatrix}
1 & - \frac{\kappa}{\sqrt{1-\kappa}} \\ 0 & \frac{1}{\sqrt{1-\kappa^2}} 
\end{pmatrix} \begin{pmatrix}
X_\mu \\ X_\mu,
\end{pmatrix},
\end{align}
where $B_\mu$ and $X_\mu$ are the physical fields of the SM and the new symmetry, respectively. The relevant covariant derivative is
\begin{align}
D_\mu = \partial_\mu +\cdots+ ig_1 Y B_\mu + i(\tilde{g} Y + g_X q_X) X_\mu,
\end{align}
where $Y$ and $g_1$ are the hypercharge and its gauge coupling,
$q_X$ and $g_X$ are the $U(1)'$ charge and its gauge coupling, and $\tilde{g}$ is the mixed gauge coupling between the groups. %The "$\cdots$" represents other terms from the SM. 
The $U(1)'$ group is spontaneously broken by the SM Higgs $H$ and introducing a new Higgs field $\chi$ with vacuum expectation values
\begin{align}
\langle H \rangle=\frac{1}{\sqrt{2}}\begin{pmatrix}
0 \\ v
\end{pmatrix} \quad \text{and} \quad \langle \chi \rangle = \frac{v'}{\sqrt{2}},
\end{align} 
respectively.
The neutral gauge boson mass eigenstates are obtained through the rotation,
\begin{align}
\begin{pmatrix}
B^\mu \\ W_3^\mu \\ X^\mu
\end{pmatrix} = 
\begin{pmatrix}
c_{\theta_W} & -s_{\theta_W} c_{\theta'} & s_{\theta_W} s_{\theta'} \\
s_{\theta_W} & c_{\theta_W} c_{\theta'} & -c_{\theta_W} s_{\theta'} \\
0 & s_{\theta'} & c_{\theta'}
\end{pmatrix}
\begin{pmatrix}
A^\mu \\ Z^\mu \\ Z'^\mu
\end{pmatrix},
\end{align}
where we have used $c_{\theta}=\cos\theta$ and $s_{\theta}=\sin\theta$, with $\theta_W$ is the weak mixing angle in the SM and $\theta'$ is the mixing angle between $Z$ and $Z'$.
After the spontaneous symmetry breaking, mass terms \citep{Cederkall:2025bka}
\begin{align}
m_Z^2 \approx \frac{v^2g_Z^2}{4}\quad \text{and} \quad m_{Z'}^2 \approx (g_{X} q_{X} v')^2
\end{align}
are induced that correspond to the neutral $Z$ and the additional vector boson $Z'$.
For $g_X$ in the order between $\mathcal{O}(10^{-4})$ and $\mathcal{O}(10^{-5})$, the mass $m_X$ can be of order $\mathcal{O}(10)$ MeV if the $v'$ is of order between  $\mathcal{O}(10^2)$ GeV and $\mathcal{O}(10^3)$ GeV. The scenario is then potentially offering a candidate to address the anomaly found in the Atomki experiment.

The interaction of this massive vector with the SM fermions is given by the gauge current
\begin{align}
J^\mu_{Z'} = \sum_{f} \overline{f} \gamma^\mu (C_{f}^{L} P_L + C_{f}^{R} P_R ) f, 
\end{align}
where $f$ denotes the fermion field, $P_{L,R}=(1\mp\gamma^5)/2$ the projection operator, and $C_{f,(L,R)}$ the chiral coefficient. We can further define the vector ($V$) and axial-vector ($A$) coefficients as
\begin{align}
C_{f}^{V} = \frac{1}{2} \left(C_{f}^{L} + C_{f}^{R}\right) \quad \text{and} \quad C_{f}^{A} = \frac{1}{2} \left(C_{f}^{L} - C_{f}^{R} \right),
\end{align}
respectively. The current then can be written as the vector and axial-vector couplings as
\begin{align}
J^\mu_{Z'} = \sum_{f} \overline{f} \gamma^\mu (C_{f}^{V} + C_{f}^{A} \gamma^5 ) f.
\end{align}
Note that both $C_f^V$ and $C_f^A$ can take either positive or negative values. This implies that the interaction exhibits $V\pm A$ couplings to the fermion $f$.
Furthermore, related to this form, the proton and neutron vector couplings are
\begin{align}
    C_p^V = 2C_u^V + C_d^V, \qquad C_n^V = C_u^V + 2C_d^V,
\end{align}
respectively. 
%Meanwhile, the axial-vector case for the proton and the neutron satisfies \citep{Kozaczuk:2016nma}
%\begin{align}
%    C_{(p,n)}^A = \sum_q \Delta_q^{(p,n)} c_q^A, 
%\end{align}
%where the values of spin-fraction for light quarks are \citep{Bishara:2017pfq}
%\begin{align}
%\begin{split}
%    &\Delta_u^p = \Delta_d^n = 0.897(27), \\
%    &\Delta_d^p=\Delta_u^n=-0.376(27), \\
%    &\Delta_s^p =\Delta_s^n=-0.031(5),
%\end{split}
%\end{align}
%and the factors for heavy quarks are small and can be safely neglected.
 
In the CE$\nu$NS process, the neutral current from the SM can be influenced by the extra gauge boson from $U(1)'$ symmetry.
Contribution coming from the additional $Z'$ boson with mass $m_{Z'}$ can be obtained by 
\begin{align} 
\begin{split}
    \left[\frac{d\sigma_{\nu_\ell}}{dT_\text{nr}}\right]_{\text{SM}+X_{17}} =  \left[1 - \frac{\sqrt{2} C_\text{eff}^{\nu_\ell}}{G_F (2 m_\mathcal{N} T_\text{nr} + m_{Z'}^2)}\right]^2 \left[\frac{d\sigma_{\nu_\ell}}{dT_\text{nr}}\right]_{\text{SM}},
\end{split}
\label{eq:x17_cevns}
\end{align}
where the effective coefficient has been defined as
\begin{align}
C_\text{eff}^{\nu_\ell} = \frac{C_{\nu_\ell}^V (Z C_{p}^V + N C_{n}^V) }{N-(1-4\sin^2\theta_W)Z}.
\end{align}
We interpret the associated neutral gauge boson $Z'$ as the $X_{17}$ particle, suggested by Atomki anomalies.
%Note that this contribution is from the vector part. 
The axial-vector interaction is generally suppressed, so that the contribution is much smaller compared to the vector interaction. For an even-even nucleus, we even have the axial-vector form factor equal to zero. 
Particularly for xenon, only $^{129}$Xe and $^{131}$Xe of their stable isotopes have non-zero spin ground states and potentially generate axial-current events. However, these are suppressed by a factor of $\sim 10^{5}$, meaning the expected SM axial-vector event rate in a multi-ton detector is effectively a fraction of a single event per decade.
With this approach, we focus on the vector part of the interaction induced by $^8$B solar neutrinos at xenon-based direct detection experiments.
The solar neutrino survival probabilities enable one to discriminate between different flavor effects. Hence, flavor-dependent limits of the effective couplings become important when one compares the bounds with the flavor-specific experiments.

\section{Data Analysis Details}
\label{section:data}

\subsection{Differential event rates}
We now estimate the differential number of events as a function of the nuclear recoil energy. We can obtain it by convolving the neutrino flux with the differential cross section as
\begin{align}
	\frac{dR}{dT_\text{nr}} = \mathcal{E} N_t \mathcal{A}(T_\text{nr})
    \int_{E_{\nu}^\text{min}}^{E_{\nu}^\text{max}} dE_\nu \frac{d  \Phi_{\nu_{\ell}}^{i}}{dE_{\nu}}  \frac{d\sigma_{\nu_\ell}}{dT_\text{nr}},
\end{align}
where $\mathcal{E}$ denotes the total experimental exposure, while $N_t$ represents  the quantity of target nuclei per unit mass within the detector. The energy-dependent detection efficiency is parameterized by $\mathcal{A}(T_\text{nr})$, and $d \Phi_{\nu_{\ell}}^{i}/dE_{\nu}$ defines the solar neutrino flux per cm$^2$ per second. Furthermore, the integral spans the accessible energy spectrum, bounded by the minimum neutrino energy $E_{\nu}^\text{min}$ and the maximum neutrino energy $E_{\nu}^\text{max}$. This lower energy threshold is determined via the following relation:
\begin{align}
	E_{\nu}^{min} = \frac{T_\text{nr}}{2}\left(1+\sqrt{1+\frac{2m_\mathcal{N}}{T_\text{nr}}} \right),
\end{align}
conversely, the maximum neutrino energy $E_{\nu}^\text{max}$ is taken as the end-point of the examined solar neutrino flux spectrum. 

For the theoretical flux profiles, we rely on the predictions formulated by the standard solar model (SSM) \citep{Bahcall:1989ks, Vinyoles:2016djt}. Recognizing that neutrinos undergo continuous flavor oscillations during their transit from the solar core to terrestrial facilities, the incident flux manifests as a coherent superposition of $\nu_e$, $\nu_\mu$, and $\nu_\tau$ states at the detector. As a result, the effective differential cross section must be rigorously weighted by the corresponding flavor-survival and transition probabilities, leading to:
\begin{align}
\frac{d\sigma}{dT_\text{nr}} = P_{ee} \frac{d\sigma_{\nu_e}}{dT_\text{nr}} + \sum_{\ell=\mu,\tau} P_{e\ell} \frac{d\sigma_{\nu_\ell}}{dT_\text{nr}}.
\end{align}
For the case of electron to muon flavor, the survival probability is
\begin{align}
P_{e\mu} = (1-P_{ee})\cos^2\vartheta_{23}
\end{align}
and for the electron to tau flavor
\begin{align}
P_{e\tau} = (1-P_{ee}) \sin^2\vartheta_{23}.
\end{align}
For the case electron flavor does not change, we consider \citep{Maltoni:2015kca}
\begin{align}
\begin{split}
P_{ee} =& \cos^2\vartheta_{13} \cos^2\vartheta_{13}^m \left[\frac{1+\cos(2\vartheta_{12}) \cos(2\vartheta_{12}^m)}{2}\right] 
 + \sin^2\vartheta_{13} \sin^2\vartheta_{12}^m.
\end{split}
\end{align}
The label $m$ denotes the matter effect, with $\cos(2\theta_{12}^m)$ being the matter angle. We consider the day-night asymmetry due to the
Earth matter effect in the calculation of the survival probabilities.  We take the neutrino oscillation parameters with normal ordering from the three-neutrino oscillation of NuFit-6.0 \citep{Esteban:2024eli}. In this work, we incorporate the day–night asymmetry arising from Earth matter effects when evaluating these probabilities. 
\begin{figure}[h!]
	\centering	
	\includegraphics[scale=0.48]{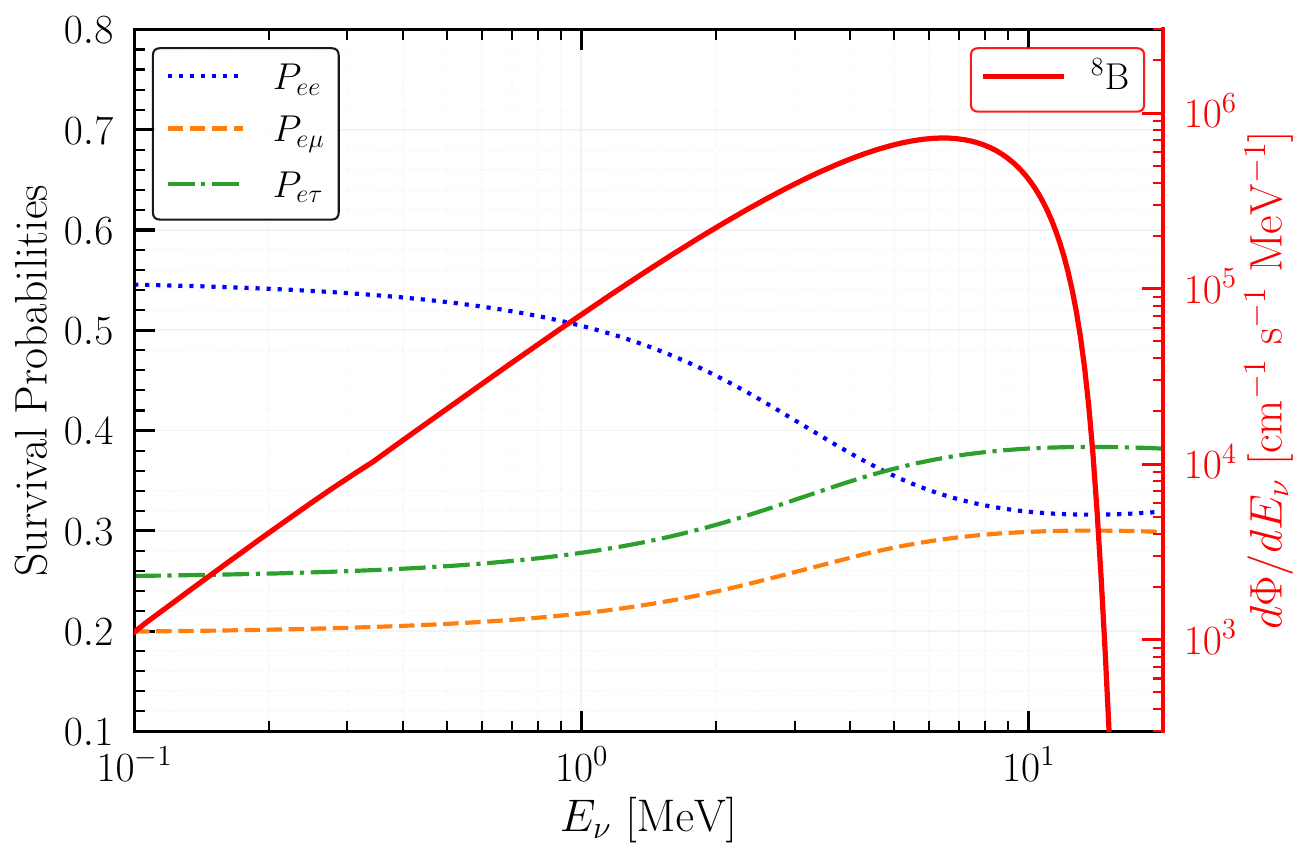}
	\caption{Averaged neutrino oscillation probabilities considering day/night asymmetry in the calculation of matter effect. The $^8$B solar neutrino flux is overlaid with the scale indicated on the right axis.}
	\label{fig:avg_prob}
\end{figure}
We provide in Fig.~\ref{fig:avg_prob} the averaged probabilities used in this work, overlaid with the solar neutrino flux from $^8$B. The effect of day-night asymmetry is considered in this work since the $^8$B solar neutrinos' energy ends at around $E_\nu\simeq 16$ MeV, which is influenced by neutrino oscillation.

\subsection{Converting nuclear recoil}
As an incoming particle interacts within the liquid phase of a direct detection experiment, two distinct signals are produced. These are a prompt scintillation signal (S1) and a delayed ionization signal (S2). The S1 signal originates from the prompt scintillation light emitted by the recoiling target atoms. The S2 signal is produced when the ionized electrons, freed during the interaction, are drifted upward by an applied electric field in the gas phase.
There, they generate secondary scintillation light via electroluminescence. The combined measurement of S1 and S2 enables accurate event reconstruction and powerful background discrimination.

The obtained data from direct detection experiments are generally given by either the number of electrons or number of photoelectrons. The XENONnT reported combination of the S1 and S2 signals, categorized as SR0 and SR1 datasets \citep{XENON:2024ijk}. The PandaX-4T provides their signal as the unpaired S2 (US2), as the available information in the experimental paper does not allow for a consistent reproduction of the results of the paired data \citep{PandaX:2024muv}. The LZ data provide S1 (paired prompt scintillation) and S2 (delayed electroluminescence) \citep{LZ:2025igz}. In this case, we need to translate the nuclear recoil energy into the relevant quantity depending on the experimental measurement. 

The differential event rate is then expressed via a change of variables
\begin{align}
\frac{dR}{dn} = \frac{dR}{dT_\text{nr}} \frac{dT_\text{nr}}{dn},
\end{align}
where $n$ might be the number of photoelectrons $n_\text{PE}$ for XENONnT, the number of electrons $n_{e^-}$ in PandaX-4T, and the number of photons detected $n_\text{phd}$ in LZ. For each experimental signal, the translations between nuclear recoil energy and these signals are
\begin{align}
n_\text{PE} =& T_\text{nr} Q_Y(T_\text{nr}) g_2^\text{XENONnT},\\
n_{e^-}  =& T_\text{nr} Q_Y(T_\text{nr}), \\
n_\text{phd} =& T_\text{nr} Q_Y(T_\text{nr}) g_2^\text{LZ},
\end{align}
where $g_2^\text{XENONnT}=16.9$ PE/electron and $g_2^\text{LZ}=34.0$ phd/electron. The factor $Q_Y(T_\text{nr})$ represents the charge yield, which is taken from Refs~\citep{PandaX:2024muv,LZ:2025igz,Sarkis:2026lds,XENON:2024xgd}.
Using this prescription, we consider the region of interest (ROI) $120<n_\text{PE} <500$ photoelectrons for XENONnT, $4<n_{e^-}<8$ for PandaX-4T, and $155<n_\text{phd}<645$ for LUX-ZEPLIN.

%%%%%%%%%%%%%%%%
\subsection{Predicted event rates}
We can calculate the number of predicted events per bin by
\begin{align}
R_i = \int_i \frac{dR}{dn} dn.
\end{align}
Since no information about them is provided in the experimental paper, the calculations do not account for resolution effects. However, even without smearing, we are able to reproduce reasonably well the predicted event rates. 
In the calculations, it is preferred to include correction factors that can be seen as effective efficiencies in order to match the predicted events in each bin with the best-fit spectra presented in experimental publications.
Hence, the total predicted number of events for each bin is obtained as
%\begin{widetext}
\begin{align}
N^\text{pre}_i = R_i(1+\alpha) + \sum_k B_i^k (1+\beta_k),
\end{align}
%\end{widetext}
where $B_i^k$ denotes the background components from each experiment.
The XENONnT backgrounds are from accidental coincidence (AC), neutron-related (neutron), and electron-recoil (ER), the PandaX-4T from cathode electron (CE) and micro-discharge (MD), while the LZ only from AC. The nuisance parameter $\alpha$ is from the neutrino flux, and $\beta_k$ from each background component.

\begin{figure*}[htp!]
	\centering	
	\includegraphics[scale=0.345]{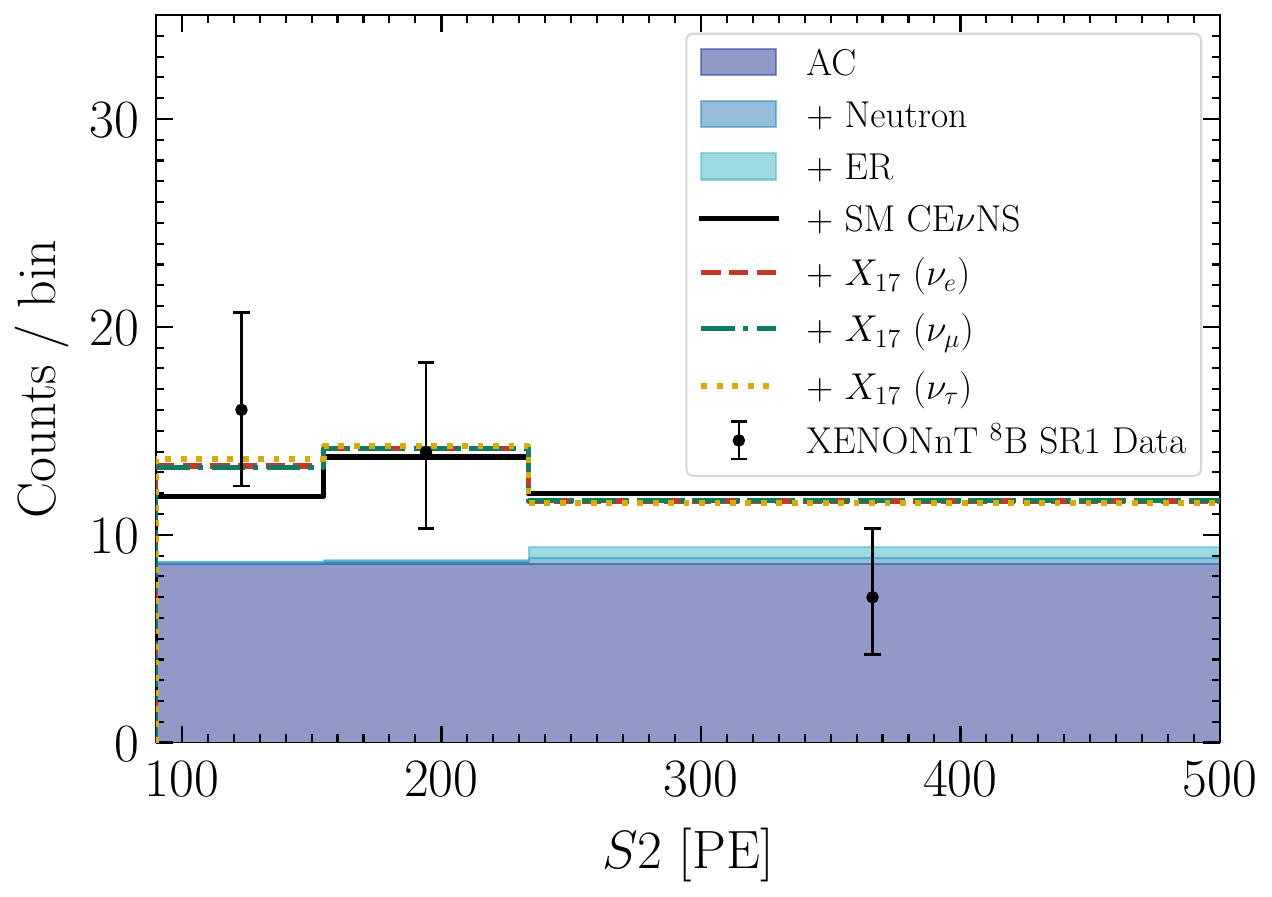} 
	\includegraphics[scale=0.345]{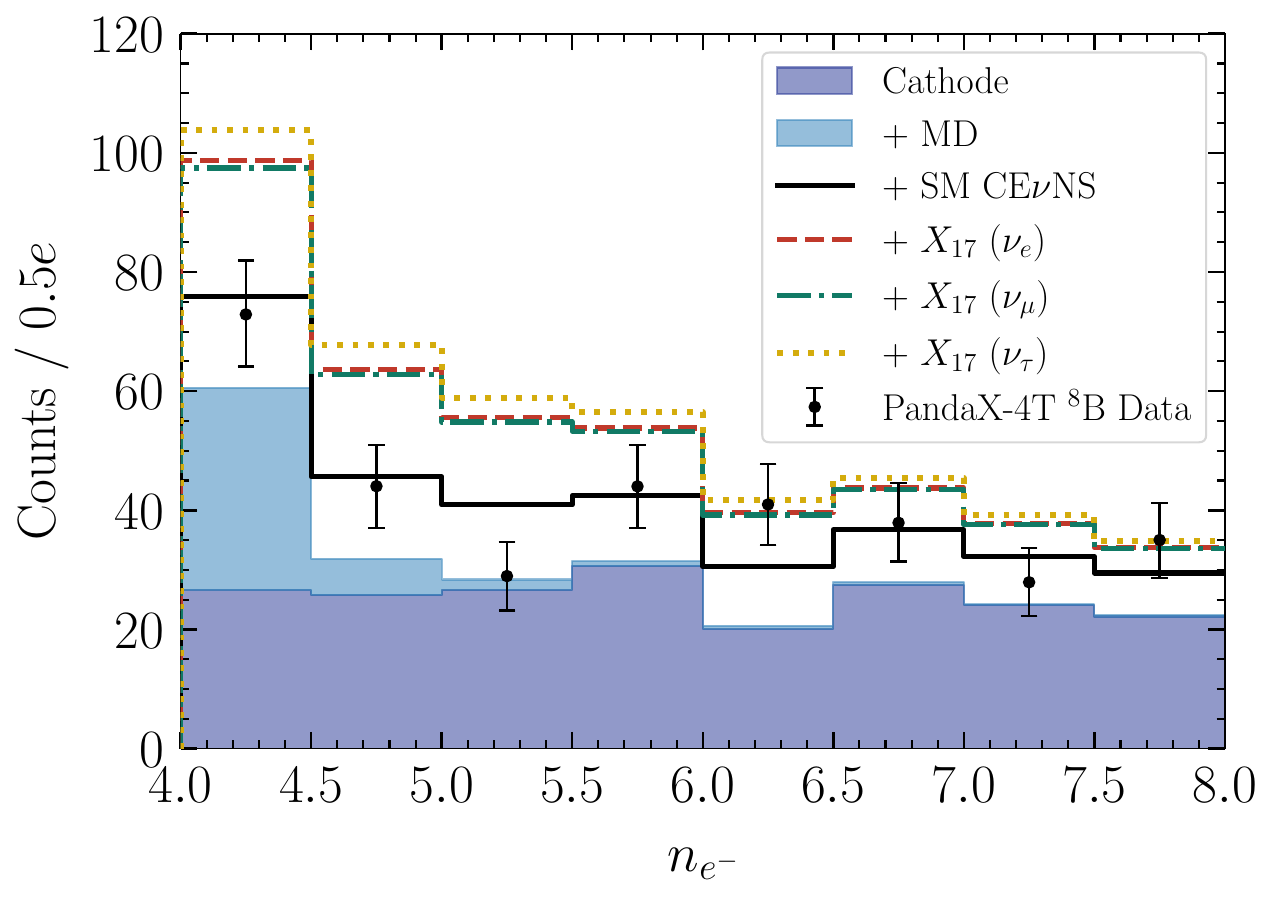}
	\\
	(a) \hspace{7.5cm} (b)
	\\
	\includegraphics[scale=0.345]{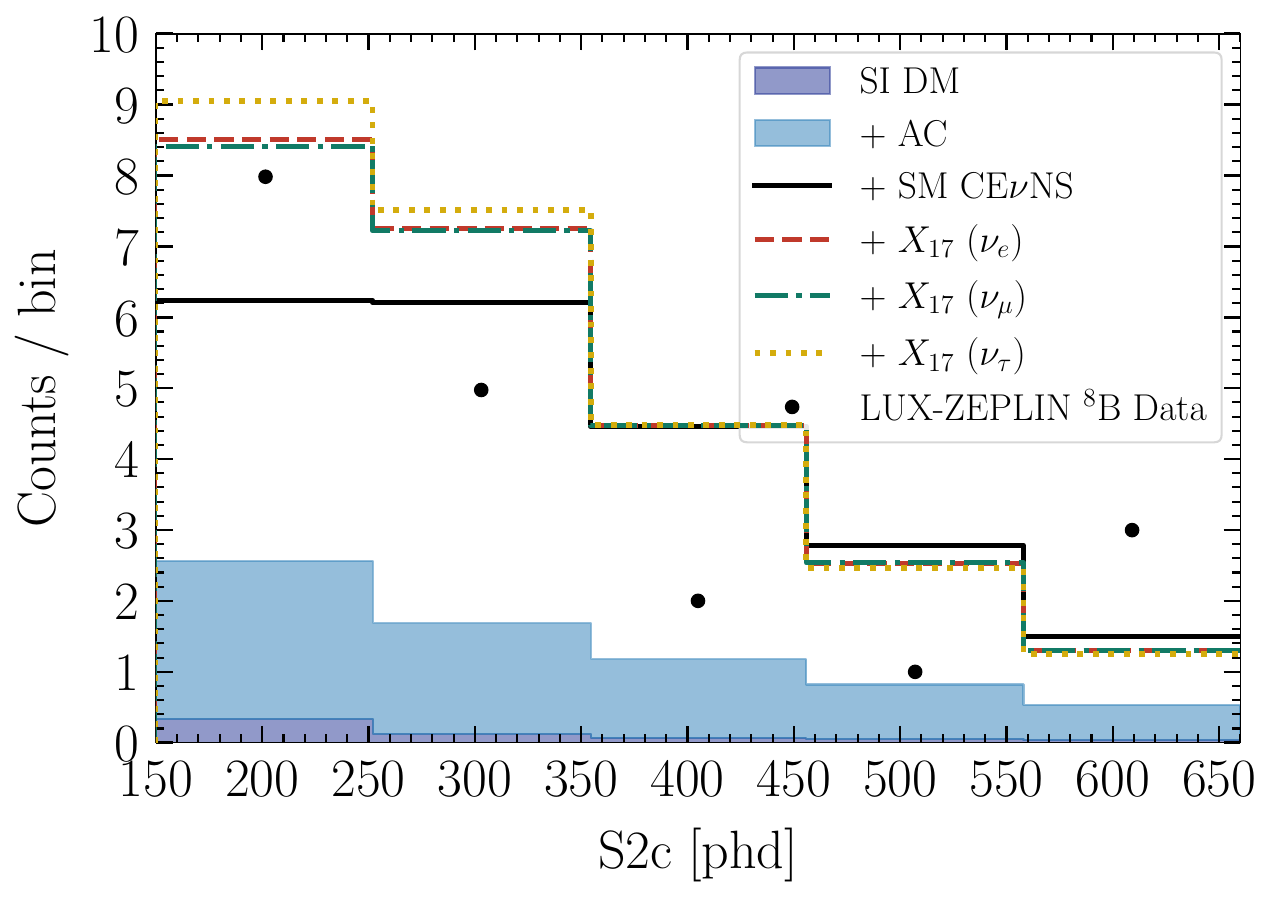}
	\\
	(c)
	\\
	\caption{Predicted event rates from (a) XENONnT (SR1), (b) PandaX-4T, and (c) LUX-ZEPLIN. Filled histograms denote the stacked background contributions from each experiment, whereas the outlined histograms represent the predicted CE$\nu$NS signals, including the contributions from $X_{17}$. The gauge-boson mass and the effective coupling values are set to $m_{Z'} = 16.7$~MeV and $C_{\text{eff}}^{\nu\ell} = 1.2 \times 10^{-8}$, respectively.}
	%(SM) CE$\nu$NS and the new physics contributions from the $X_{17}$ particle
	\label{fig:event_rate}
\end{figure*}

\begin{figure*}[ht!]
	\centering	
	\includegraphics[scale=0.39]{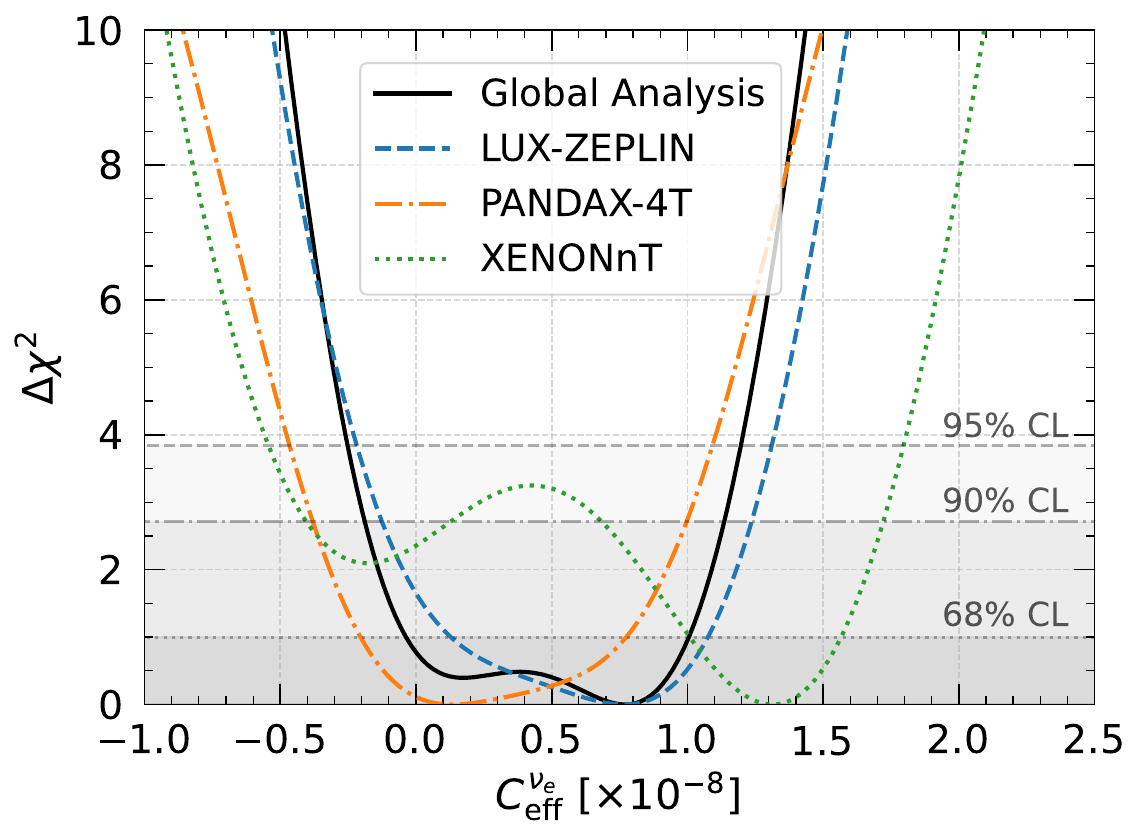} 
	\includegraphics[scale=0.39]{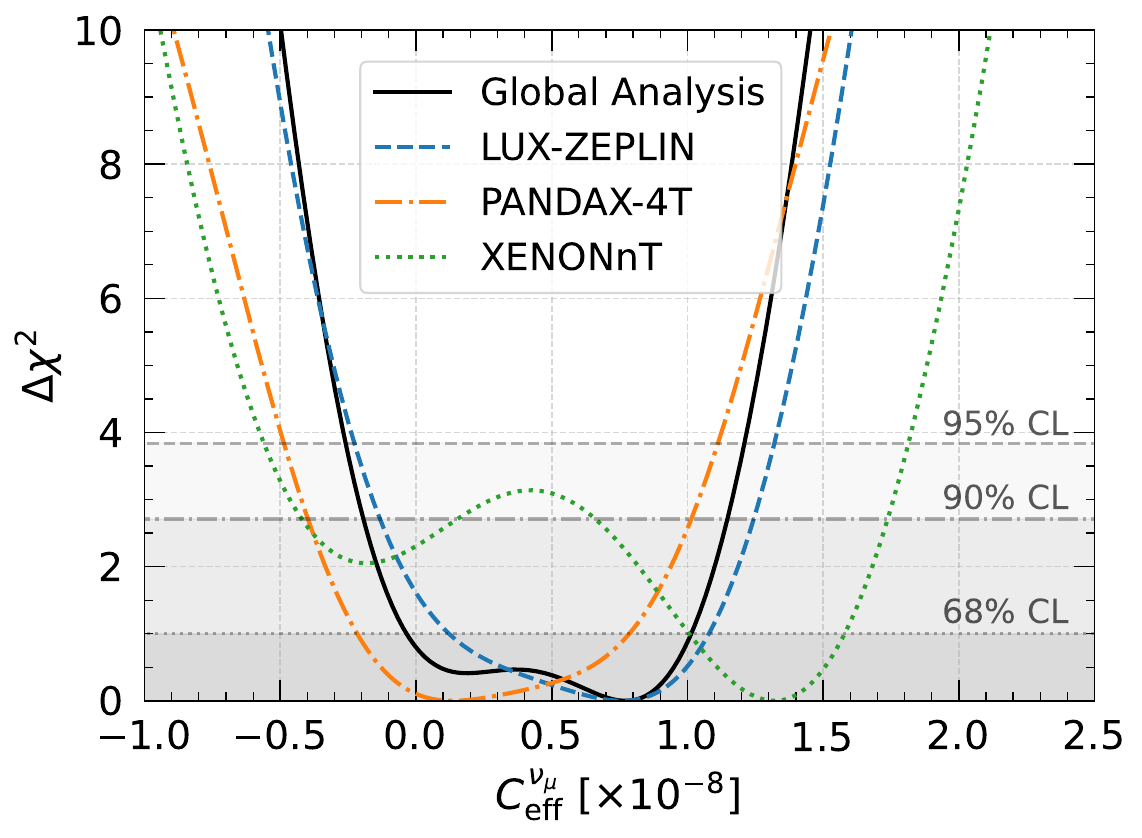}
	\\
	(a) \hspace{7.5cm} (b)
	\\
	\includegraphics[scale=0.39]{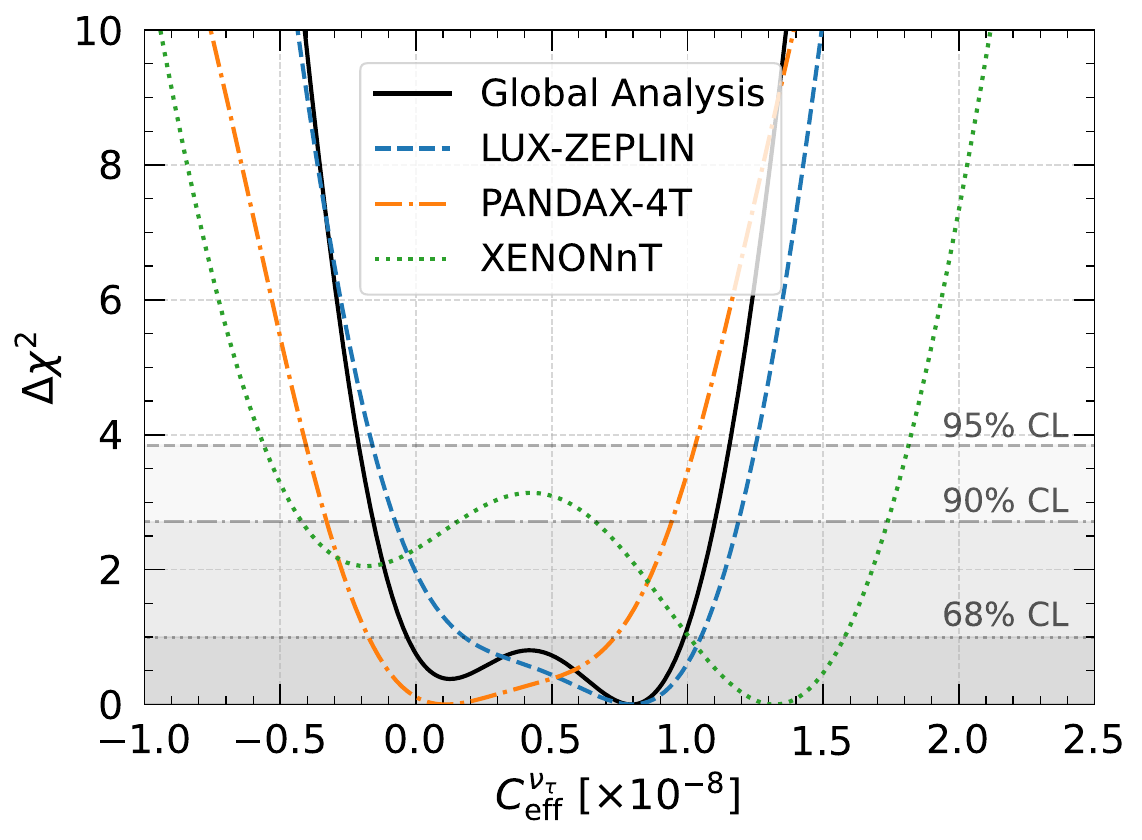}
	\\
	(c)
	\\
	\caption{The marginal $\Delta\chi^2$ profiles for the effective vector couplings (a) $C_\text{eff}^{\nu_e}$, (b) $C_\text{eff}^{\nu_\mu}$, and (c) $C_\text{eff}^{\nu_\tau}$, obtained from the analysis of XENONnT, PandaX-4T, and LUX-ZEPLIN datasets, along with their global combination, assuming $m_{Z'}=16.7$~MeV. In each panel, the profiles are marginalized by minimizing the $\Delta\chi^2$with respect to the other remainingŁ effective couplings.}    
	\label{fig:limit_1dof}
\end{figure*}

\begin{figure}[ht!]
	\centering	
	\includegraphics[scale=0.43]{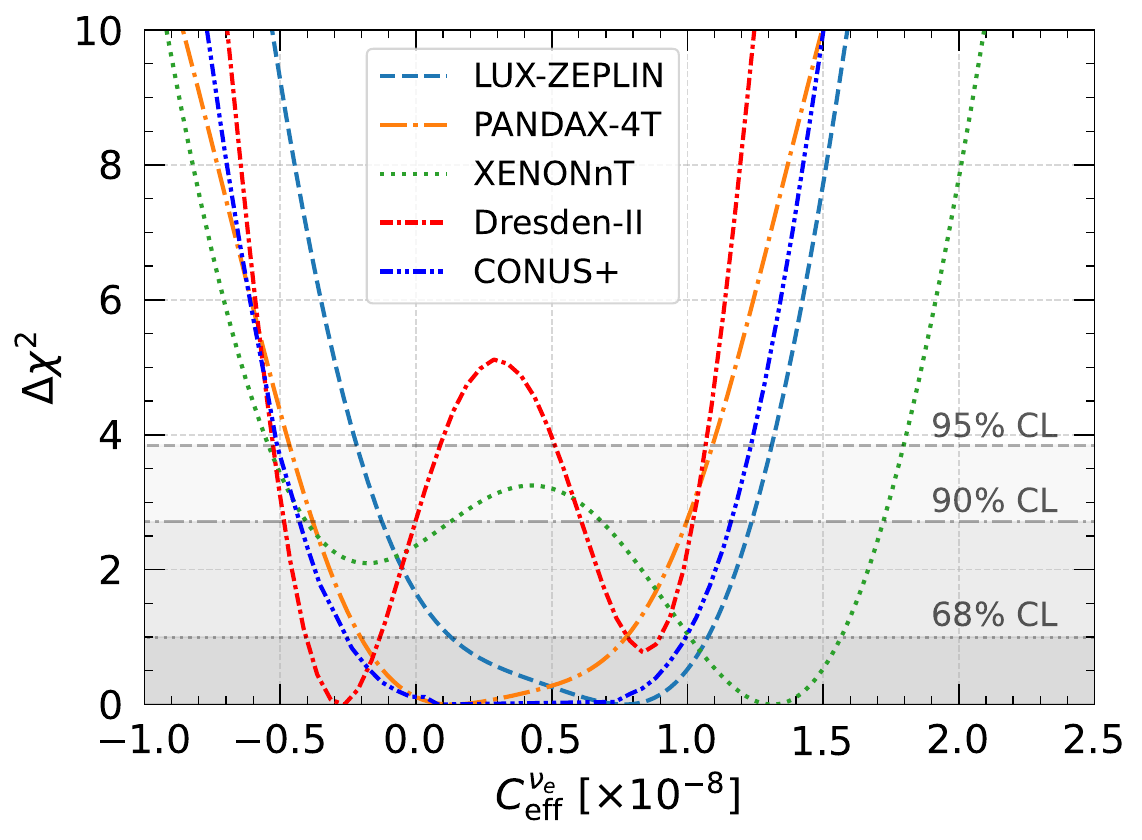} 
    \caption{The $\Delta\chi^2$ profiles for the effective vector coupling $C_\text{eff}^{\nu_e}$ from our analysis of solar neutrino datasets (XENONnT, PandaX-4T, LUX-ZEPLIN), compared with the individual reactor constraints from Dresden-II and CONUS+ (incorporating QF uncertainties) adapted from Ref.~\cite{Rathsman:2026smv}. For our bounds, the profiles are marginalized by minimizing the $\Delta\chi^2$ with respect to the other remaining effective couplings. All limits are derived for $m_{Z'}=16.7$~MeV.}
	\label{fig:limit_1dof_comp}
\end{figure}

\subsection{The $\chi^2$ function}
The statistical analysis is carried out by using the $\chi^2$ function that satisfies \cite{Baker:1983tu,Fogli:2002pt}
%\begin{widetext}
\begin{eqnarray}
	\begin{split}
	\chi^2 = \mathrm{min}_{\alpha,\beta_k} \Bigg\{& 2 \left[\sum_i N^\text{pre}_i - N_i^\text{exp} + N_i^\text{exp} \ln\left(\frac{N_i^\text{exp}}{N_i^\text{pre}}\right) \right] + \left(\frac{\alpha}{\sigma_\alpha}\right)^2 + \sum_k \left(\frac{\beta_k}{\sigma_{\beta_k}}\right)^2 \Bigg\}.
	\end{split}
\end{eqnarray} 
%\end{widetext}
The $\chi^2$-function is minimized according to the nuisance parameters. The $\sigma_\alpha=12\%$ is from the $^8$B solar neutrino flux \citep{Baxter:2021pqo}. Other backgrounds are from the considered experiments. For XENONnT, we use the different contributions from AC, neutron, and ER backgrounds with relative uncertainties $\sigma_\text{AC}=9\%$, $\sigma_\text{neutron}=54\%$, and $\sigma_\text{ER}=100\%$ for SR0 and $\sigma_\text{AC}=5.8\%$, $\sigma_\text{neutron}=58\%$, and $\sigma_\text{ER}=100\%$ for SR1, as well as $5\%$ in the signal prediction due to the fiducial volume \citep{XENON:2024ijk}. For PandaX-4T, the uncertainties are $\sigma_\text{CE}=31\%$ and $\sigma_\text{MD}=23\%$, and also a $22\%$ uncertainty in signal prediction due to data selection and interaction modeling \citep{PandaX:2024muv}.
For the LUX-ZEPLIN, the uncertainty is $\sigma_\text{AC}=30\%$ \citep{LZ:2025igz}. In performing the analysis, we implement marginalization over all flavor components of the effective couplings.

\begin{table}[ht!]
\caption{Upper limits with 1 dof on the $C_\text{eff}^{\nu_e}$, $C_\text{eff}^{\nu_\mu}$, and $C_\text{eff}^{\nu_\tau}$ obtained from XENONnT, PandaX-4T, and LUX-ZEPLIN datasets, as well as global analysis from combining these results. These limits are derived for the $m_{Z'} = 16.7$ MeV. Other limits from Dresden-II and CONUS+ for $C_\text{eff}^{\nu_\ell}$ derived in Ref.\citep{Rathsman:2026smv} are included for comparison.
}
\begin{center}
\resizebox{\textwidth}{!}{%
\begin{tabular}{l  l   c  c  c} 
 \hline
 \hline
 \multirow{2}{1.5cm}{Couplings ($\times 10^{-8}$)} & Experiments &  68\% CL & 90\% CL & 95\%  CL \\
  \hline
  & XENONnT & $[-0.54, 1.80]]$ & $[-0.40, 0.13]\cup[0.68, 1.72]$ & $[1.01, 1.57]$\\
  \cline{2-5}
  & PandaX-4T & $[-0.47, 1.09]$ & $[-0.38, 1.00]$ & $[-0.47, 1.09]$\\
  \cline{2-5}
  $\quad$ $C_\text{eff}^{\nu_e}$ $\quad$ & LUX-ZEPLIN & $[0.13, 1.08]$ & $[-0.13, 1.24]$ & $[-0.22, 1.31]$\\
  \cline{2-5}
  & Global Analysis & $[-0.04, 1.01]$ & $[-0.19, 1.14]$ & $[-0.25, 1.20]$\\
  \cline{2-5}
  & Dresden-II \cite{Rathsman:2026smv} & $[-0.40, -0.13]\cup[0.78, 0.9]$ & $ [-0.49, -0.003]\cup  [0.61, 1.02]$ & $[-0.53, 0.09]\cup[0.51, 1.07]$\\
  \cline{2-5}
  & CONUS+ \cite{Rathsman:2026smv} & $[-0.26, 0.99]$ & $[-0.43, 1.16]$ & $[-0.51, 1.23]$\\
  \hline
  & XENONnT & $[-0.56, 1.81]$ & $ [-0.42, 0.15]\cup[0.66, 1.74]$ & $[1.01, 1.58]$\\
  \cline{2-5}
  $\quad$ $C_\text{eff}^{\nu_\mu}$ $\quad$
  & PandaX-4T & $[-0.22, 0.79]$ & $[-0.4, 1.01]$ & $[-0.49, 1.11]$\\
  \cline{2-5}
  & LUX-ZEPLIN & $[0.12, 1.08]$ & $ [-0.13, 1.24]$ & $[-0.23, 1.32]$\\
  \cline{2-5}
  & Global Analysis & $[-0.03, 1.01]$ & $[-0.19, 1.15]$ & $[-0.26, 1.21]$\\
  \hline
  & XENONnT & $[1.01, 1.48]$ & $[-0.27, 0.008]\cup[0.78, 1.62]$ & $[-0.42, 0.33]\cup[0.55, 1.68]$  \\
  \cline{2-5}
  $\quad$ $C_\text{eff}^{\nu_\tau}$ $\quad$  
  & PandaX-4T & $[-0.18, 0.73]$ & $[-0.33, 0.94]$ & $[-0.40, 1.03]$\\
  \cline{2-5}
  & LUX-ZEPLIN & $[0.18, 1.05]$ & $[-0.08, 1.19]$ & $[-0.16, 1.25]$\\
  \cline{2-5}
   & Global Analysis & $[-0.03, 0.99]$ & $[-0.16, 1.10]$ & $[-0.21, 1.15]$\\
  \hline
  \hline
\end{tabular}
		}
\label{tab:limit_1dof}
\end{center}
\end{table}
%%%

\section{Results}\label{sec:results}
In this section, we present constraints on the $X_{17}$ effective couplings derived from our analysis of the XENONnT, PandaX-4T, and LUX-ZEPLIN datasets. These are relevant to the nuclear recoils from the CE$\nu$NS process triggered by the $^8$B solar neutrino. To properly account for neutrino flavor dependence, we incorporate the solar neutrino survival probabilities with day-night asymmetry. By employing the $\chi^2$ formalism, we determine the allowed parameter space for the effective couplings at various confidence levels (CLs) based on the relevant degrees of freedom (dof). 
Throughout our analysis, all effective couplings are treated as free parameters, allowing us to present fully marginalized constraints.

We first plot the predicted event rates due to the $X_{17}$ particle in the CE$\nu$NS process for XENONnT, PandaX-4T, and LUX-ZEPLIN in Figs.~\ref{fig:event_rate}(a), \ref{fig:event_rate}(b), and \ref{fig:event_rate}(c), respectively. For illustrative purposes, we fix the gauge-boson mass at $m_{Z'} = 16.7$~MeV and the effective couplings at $C^{\nu_\ell}_\text{eff}=1.2\times 10^{-8}$. The new physics contributions are evaluated by incorporating the $X_{17}$ mediator into the $^8$B solar neutrino CE$\nu$NS framework. As depicted in the histograms, the inclusion of this new boson yields a notable distortion in the expected signal spectrum, producing a significant excess primarily in the lowest recoil energy bins compared to the SM baseline. Furthermore, the dominant contribution clearly arises from the $\nu_\tau$ flavor (yellow dotted lines), followed by the $\nu_e$ and $\nu_\mu$ flavors. This distinct hierarchy visually demonstrates the profound impact of the averaged neutrino survival probabilities when convoluting the $^8$B solar neutrino flux, and highlights how these low-energy regions provide the most stringent constraining power against the experimental data.

We present the 1~dof marginalized $\Delta\chi^2$ profiles for the effective vector couplings in Fig.~\ref{fig:limit_1dof}. The constraints on $C_\text{eff}^{\nu_e}$ are shown in Fig.~\ref{fig:limit_1dof}(a). As depicted, the LUX-ZEPLIN dataset provides the most stringent limit, followed closely by PandaX-4T. In contrast, the XENONnT profile exhibits a broader, degenerate structure with dual minima, resulting in relatively weaker constraints. Furthermore, we are able to perform global analysis by combining results from these datasets. Quantitatively, the 90\% CL allowed range derived from our global analysis is
\begin{equation}
-0.19\times10^{-8} \lesssim C_\text{eff}^{\nu_e} \lesssim 1.14\times 10^{-8}.
\end{equation}
Similar behaviors of $\Delta\chi^2$ are observed for $C_\text{eff}^{\nu_\mu}$ and $C_\text{eff}^{\nu_\tau}$, as illustrated in Figs.~\ref{fig:limit_1dof}(b) and \ref{fig:limit_1dof}(c), respectively. For these couplings, the 90\% CL allowed intervals from the global analysis are obtained as
\begin{align}
-0.19\times10^{-8} &\lesssim C_\text{eff}^{\nu_\mu} \lesssim 1.15\times 10^{-8}, \\
-0.16\times10^{-8} &\lesssim C_\text{eff}^{\nu_\tau} \lesssim 1.10\times 10^{-8}.
\end{align}
Furthermore, in Fig.~\ref{fig:limit_1dof_comp}, we compare our findings for $C_\text{eff}^{\nu_e}$ with the existing constraints from the Dresden-II and CONUS+ reactor experiments, adapted from Ref.~\cite{Rathsman:2026smv} with $m_{Z'}=16.7$ MeV. Our limits derived from solar neutrinos at the direct detection searches indicate a significant improvement over these bounds obtained utilizing reactor electron antineutrino sources. A comprehensive summary of our derived 68\%, 90\%, and 95\% CL limits for each dataset, including the global analysis limits, is provided in Table~\ref{tab:limit_1dof}.

\begin{figure*}[hbt!]
	\centering	
	\includegraphics[scale=0.5]{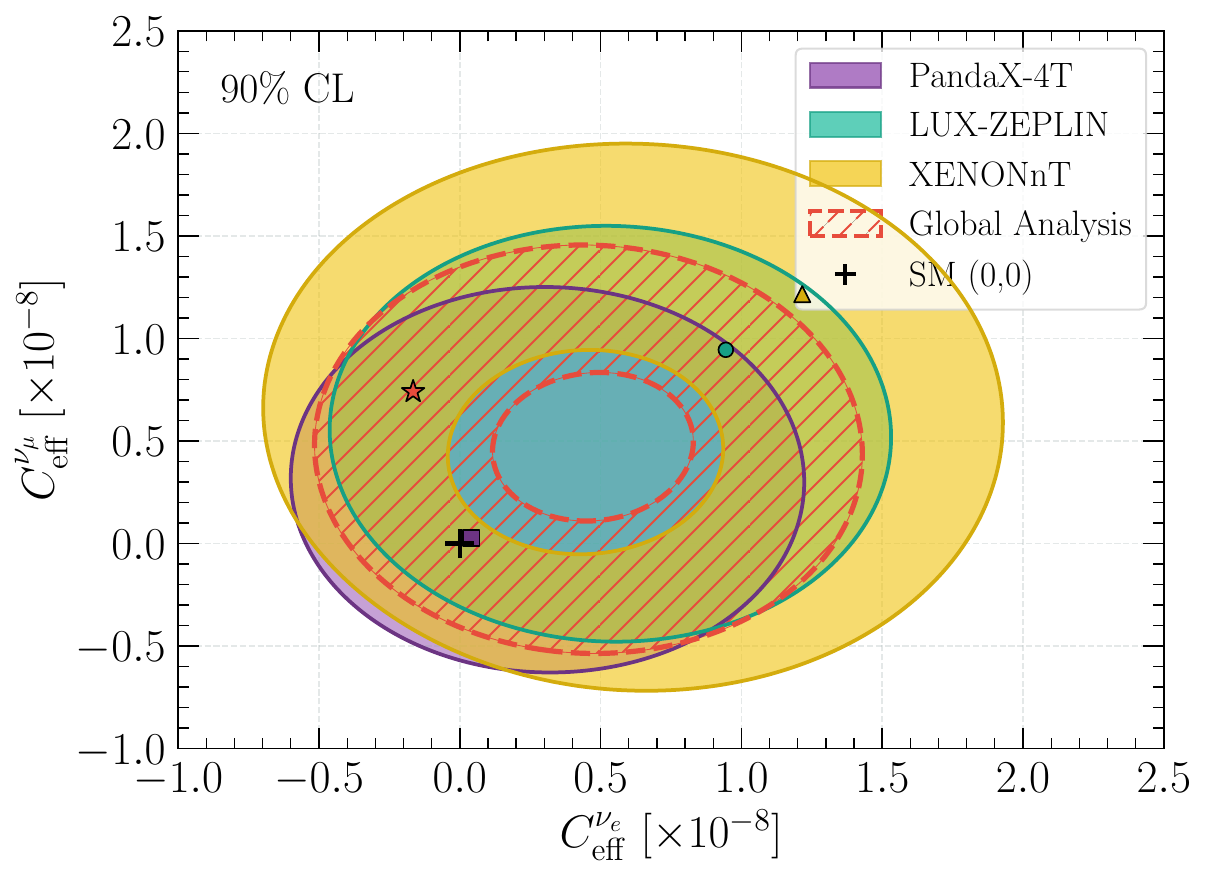} 
\caption{The 90\% CL allowed regions in the effective vector coupling parameter space ($C_\text{eff}^{\nu_e}$, $C_\text{eff}^{\nu_\mu}$) obtained from XENONnT, PandaX-4T and LUX-ZEPLIN datasets, and their global analysis. The $\Delta\chi^2$ function is minimized with respect to the effective coupling parameter not shown ($C_\text{eff}^{\nu_\tau}$). The SM prediction is denoted by the black cross. The best-fit points from XENONnT, PandaX-4T, and LUX-ZEPLIN are indicated by the yellow triangle, purple square, and cyan circle, respectively, whereas the global best fit is represented by the red star.}
	\label{fig:limit_2dof_emu}
\end{figure*}
%%%%
\begin{figure}[ht!]
	\centering	
	\includegraphics[scale=0.5]{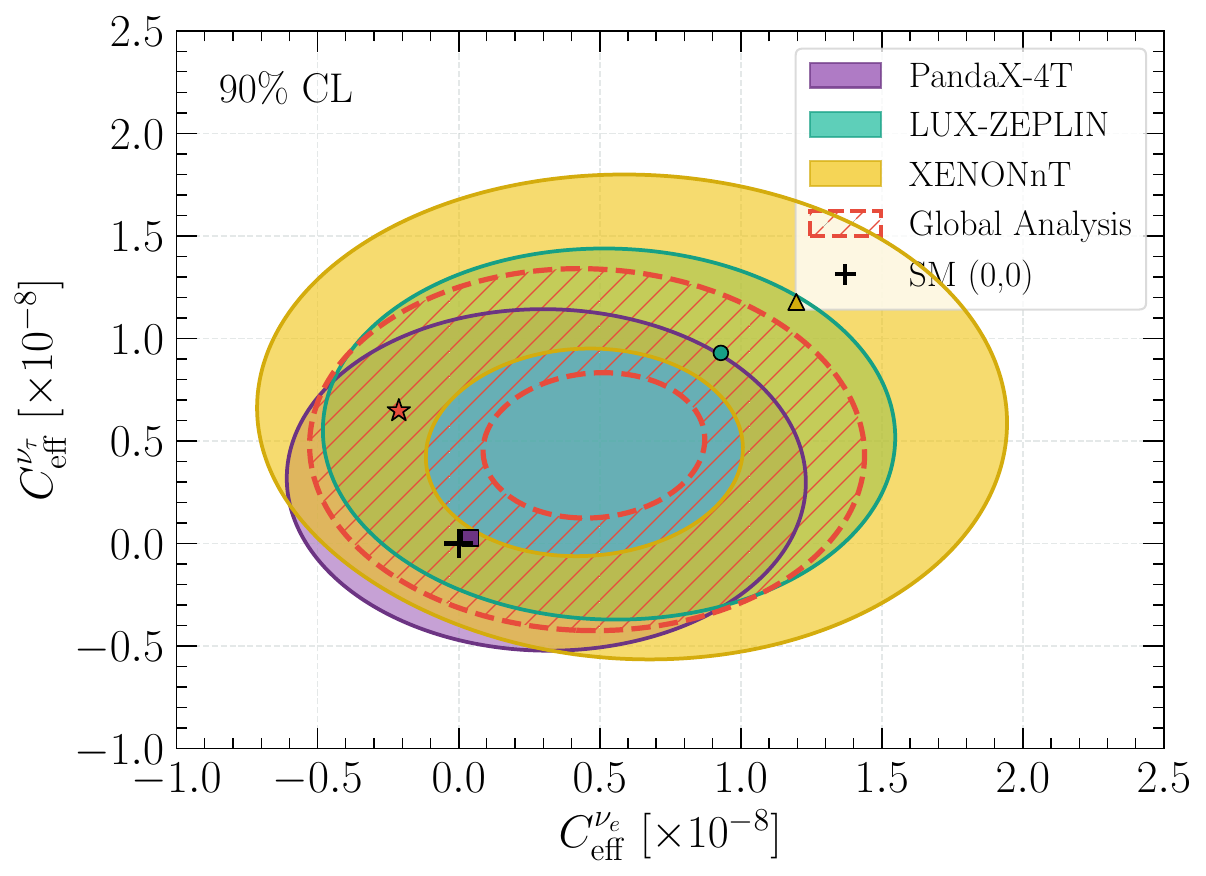}
	\caption{Same as Fig.\ref{fig:limit_2dof_emu}, but for parameter space ($C_\text{eff}^{\nu_e}$ , $C_\text{eff}^{\nu_\tau}$).}
	\label{fig:limit_2dof_etau}
\end{figure}
%%%%%
\begin{figure}[ht!]
	\centering	
	\includegraphics[scale=0.5]{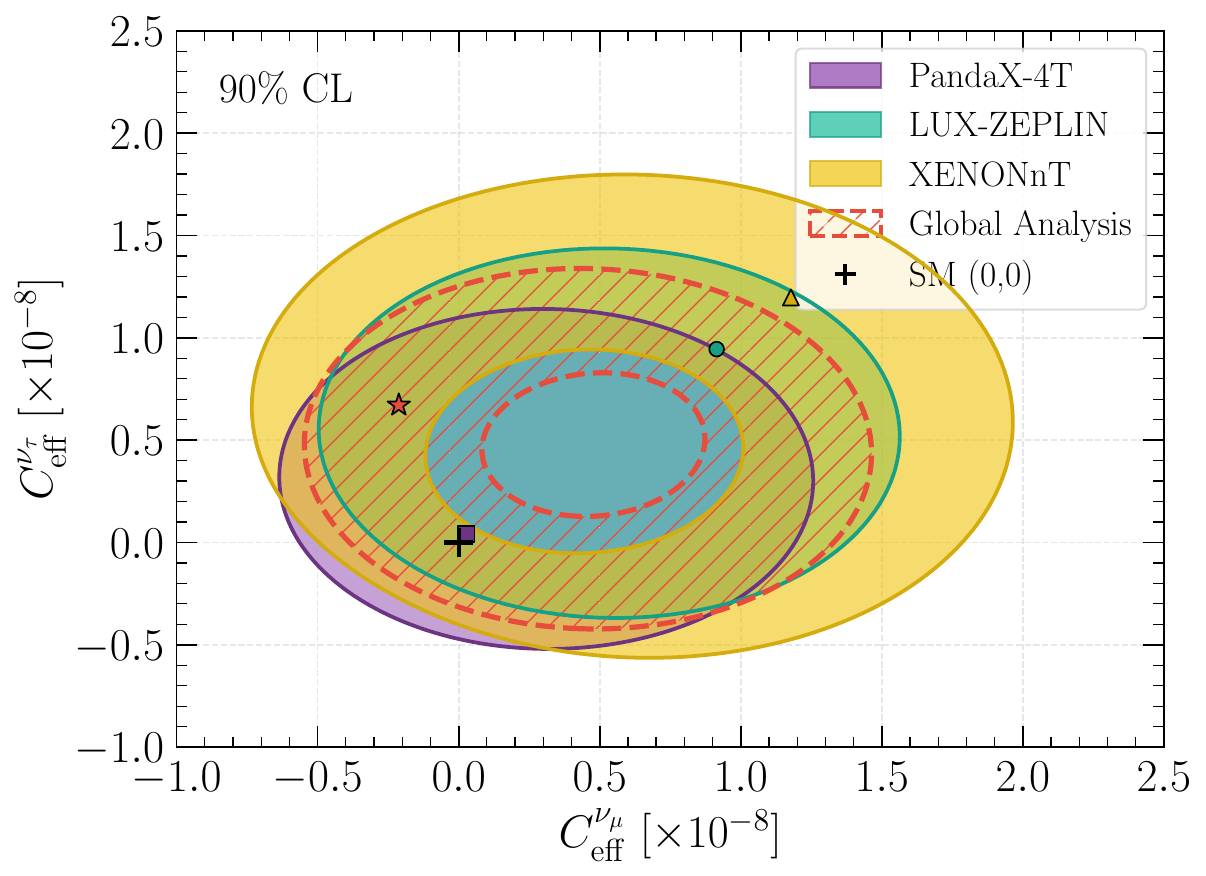}
	\caption{Same as Fig.\ref{fig:limit_2dof_emu}, but for parameter space ($C_\text{eff}^{\nu_\mu}$ , $C_\text{eff}^{\nu_\tau}$).}
	\label{fig:limit_2dof_mutau}
\end{figure}
%%%

Moving beyond the single-parameter constraints, we now explore the correlations between the effective couplings by analyzing the two-dimensional (2~dof) parameter spaces. 
We present in Fig.~\ref{fig:limit_2dof_emu} the 90\% CL allowed regions from each dataset in the ($C_\text{eff}^{\nu_e}$, $C_\text{eff}^{\nu_\mu}$) parameter space, derived by marginalizing over the effective coupling of the imperceptible $\nu_\tau$ flavor ($C_\text{eff}^{\nu_\tau}$). 
The derivation corresponds to those values that produce $\chi^2 = \chi^2_\text{min} + 4.61$. 
In this parameter space, the contours reveal a striking interplay between the datasets. 
Notably, the PandaX-4T and LUX-ZEPLIN limits share almost identical outer boundaries in terms of scale and orientation; however, their internal topologies and best-fit centers are distinctively shifted. 
PandaX-4T provides a contiguous, simply connected elliptical region with its best-fit point (purple square) resting remarkably close to the SM expectation at ($0,0$). 
In contrast, LUX-ZEPLIN exhibits an annular structure with a central void, significantly shifting its best-fit candidate (cyan circle) toward higher effective coupling values. 
XENONnT further expands on this trend, presenting a much broader annular allowed region that also completely encapsulates the zero-coupling area. Because the annular structures from LUX-ZEPLIN and XENONnT statistically favor non-zero couplings, the combined global best-fit point (red star) is visibly pulled away from the origin. Nevertheless, the robust, SM-encompassing core provided by PandaX-4T strongly anchors the global analysis (red hatched region). 
As a result, individual degeneracies are effectively broken, and the SM hypothesis is completely protected, remaining comfortably within the combined 90\% C.L. allowed parameter space.

We extend to explore the ($C_\text{eff}^{\nu_e}$, $C_\text{eff}^{\nu_\tau}$) and ($C_\text{eff}^{\nu_\mu}$, $C_\text{eff}^{\nu_\tau}$) parameter spaces, illustrated in Fig.~\ref{fig:limit_2dof_etau} and Fig.~\ref{fig:limit_2dof_mutau}, respectively. These limits are obtained by marginalizing over the effective couplings of the $\nu_\mu$ and $\nu_e$ flavors, correspondingly. The topological features—including the structural similarities between PandaX-4T and LUX-ZEPLIN, and the shifting of the best-fit centers—closely mirror those of the first case. 
It is worth mentioning that the inclusion of the $\nu_\tau$ coupling parameter is particularly significant from a physical standpoint. Due to flavor transitions during propagation from the Sun, a substantial fraction of $^8$B solar neutrinos arrive at the Earth as $\nu_\tau$. Consequently, these direct detection experiments demonstrate a remarkable sensitivity to the tau-flavor coupling of the $X_{17}$ mediator, which is distinctly reflected in the overall scale and orientation of the global contours.

%%%%%%%%%%%%%%%%%%%%%
\begin{figure}[ht!]
	\centering	
	\includegraphics[scale=0.5]{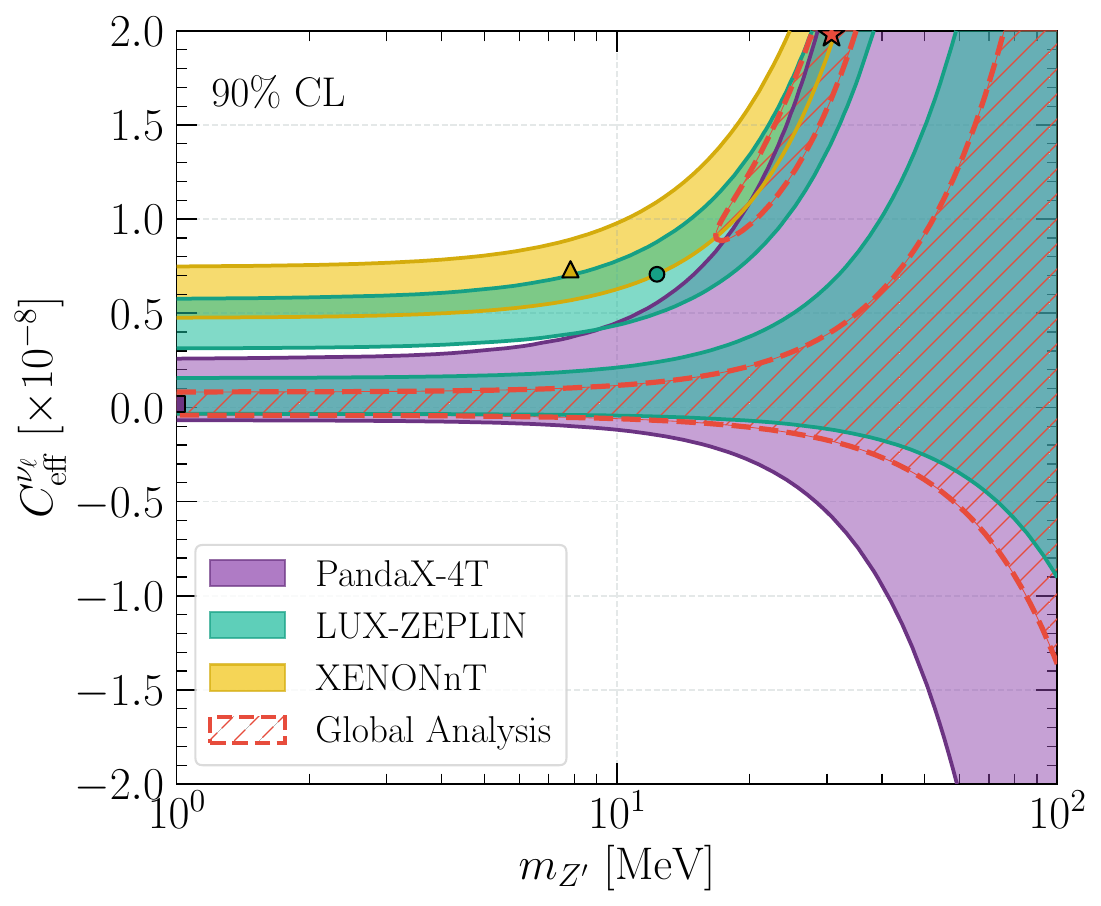}
\caption{The 90\% CL allowed regions in the flavor-independent parameter space ($m_{Z'}$, $C_\text{eff}^{\nu_\ell}$) constraints from XENONnT, PandaX-4T, and LUX-ZEPLIN datasets, along with their global analysis. The best-fit points from XENONnT, PandaX-4T, and LUX-ZEPLIN are indicated by the yellow triangle, purple square, and cyan circle, respectively, while the global best fit is represented by the red star.}
	\label{fig:limit_2dof_Ceff_mass}
\end{figure}
%%%%%%%%%%%%%%%%%%
\begin{figure}[ht!]
	\centering	
	\includegraphics[scale=0.56]{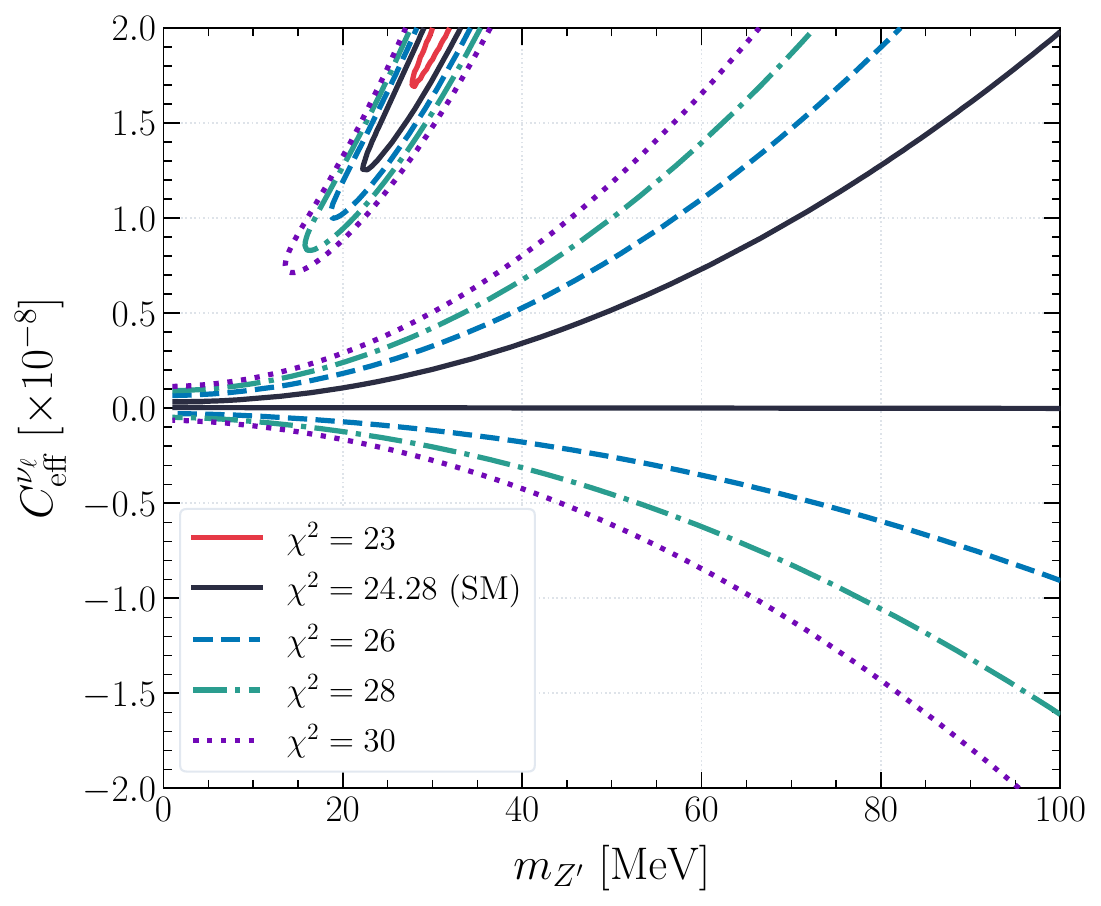}
	\caption{The $\chi^2$ contours in the ($m_{Z'}$, $C_\text{eff}^{\nu\ell}$) parameter space obtained from the global analysis of XENONnT, PandaX-4T, and LUX-ZEPLIN datasets. The solid black line indicates the SM prediction ($\chi^2 = 24.28$). The dashed and dotted curves correspond to various $\chi^2$ thresholds.}
	\label{fig:limit_2dof_Ceff_mass_chi2}
\end{figure}
%%%%%%%%%%%%%%%%%%%%%%%

In the preceding evaluations, we set the gauge-boson mass as a fixed quantity. We now relax this assumption and allow $m_{Z'}$ to vary alongside the flavor-independent effective coupling $C_\text{eff}^{\nu_\ell}$. We present the resulting 90\% CL allowed regions in the $(m_{Z'}, C_\text{eff}^{\nu_\ell})$ parameter space in Fig.~\ref{fig:limit_2dof_Ceff_mass}. Similar to the fixed-mass bounds, PandaX-4T and LUX-ZEPLIN indicate almost identical shape, particularly in the low-mass regime ($m_{Z'} \lesssim 10$~MeV) where the allowed effective coupling narrows significantly toward zero. Notably, the LUX-ZEPLIN dataset introduces a distinct topological degeneracy, splitting the allowed parameter space into two separate bands.
Meanwhile, XENONnT provides a much broader region. At higher masses, the constraints universally relax. 
When statistically combining the datasets, the global analysis (red dashed contours) isolates a preferred closed region—an ``island"—in the high-mass parameter space, specifically for $m_{Z'} \gtrsim 17$~MeV and $C_\text{eff}^{\nu_\ell} \gtrsim 0.8 \times 10^{-8}$, while still maintaining a narrow allowed band consistent with zero coupling.

To further elucidate this statistical topology, we map the absolute $\chi^2$ contours across the $(m_{Z'}, C_\text{eff}^{\nu_\ell})$ plane in Fig.~\ref{fig:limit_2dof_Ceff_mass_chi2}. The SM prediction yields $\chi^2_\text{SM} = 24.28$, denoted by the solid black curves, in accordance with the result in Ref.\citep{Rathsman:2026smv}. Intriguingly, the global best-fit (the minimum of the $\chi^2$ surface) achieves a value of $\chi^2 = 23$, which strictly localizes within the upper parameter space region ($m_{Z'} \gtrsim 15$~MeV and $C_\text{eff}^{\nu_\ell} \gtrsim 0.7 \times 10^{-8}$). This $\Delta\chi^2 \simeq 1.28$ improvement indicates a slight statistical preference for a massive $Z'$ mediator with a non-zero coupling, driving the closed contour seen in the global analysis.
We emphasize that our derived limits fundamentally depend on the explicit incorporation of the quenching factor (QF) uncertainties related to the charge yields. We note that employing different QF treatments or experimental prescriptions, such as those utilized in Ref.~\cite{Rathsman:2026smv}, may alter the contour topologies and lead to different physical interpretations.

%%%%%%%%%%%%%
\begin{figure*}[ht!]
	\centering	
	\includegraphics[scale=0.64]{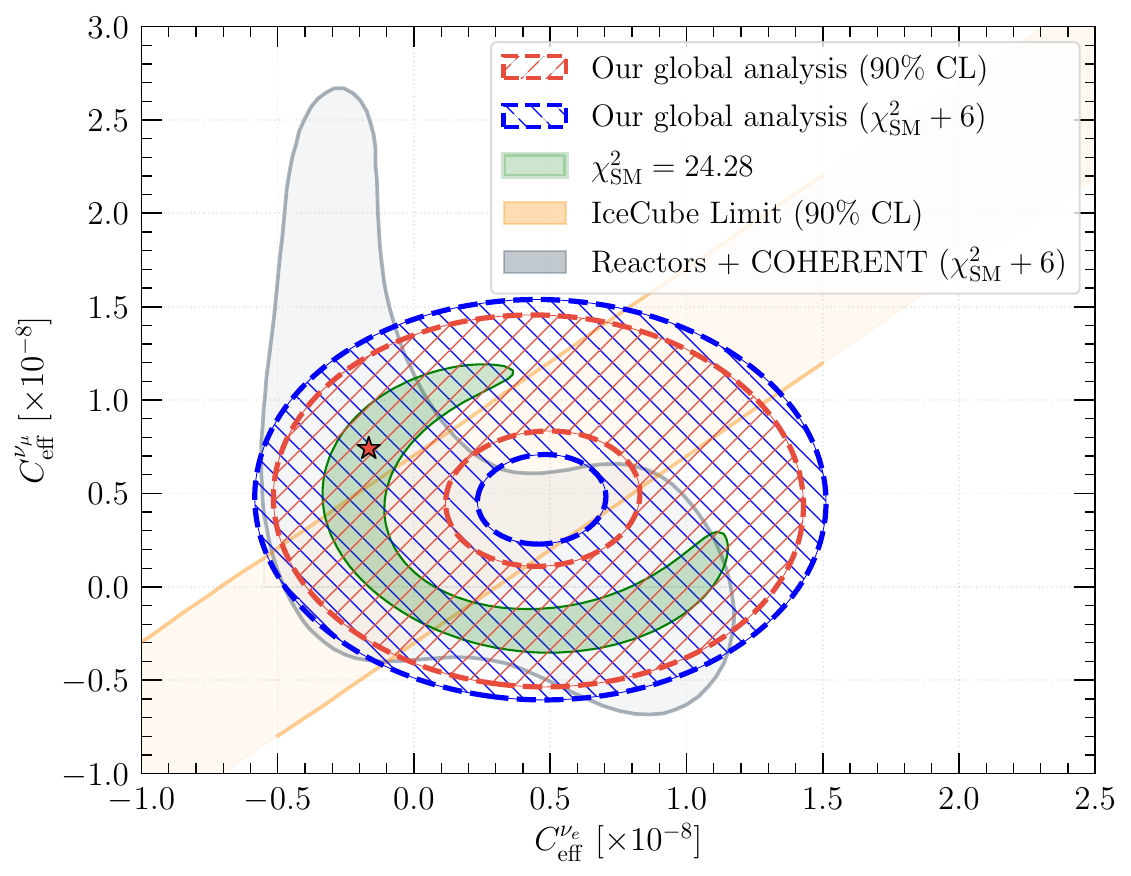}
	\caption{Comparison of the allowed regions in the ($C_\text{eff}^{\nu_e}$, $C_\text{eff}^{\nu_\mu}$) parameter space from our global analysis (showing both the 90\% CL and $\chi^2_{\text{SM}} + 6$ contours) against various constraints from the literature. The IceCube limit (90\% CL) is adapted from Ref.~\cite{Rathsman:2026hif}, while the combined reactor and COHERENT bounds ($\chi^2_{\text{SM}} + 6$, incorporating QF uncertainties) are taken from Ref.~\cite{Rathsman:2026smv}. The green region highlights the $\chi^2_{\text{SM}} = 24.28$ parameter space. The red star indicates our global best-fit point.}
	\label{fig:limit_2dof_comp}
\end{figure*}
%%%%%%
%%%%%%%%
Finally, to contextualize the strength of our bounds, we compare our global analysis with existing constraints from IceCube (90\% CL) \cite{Rathsman:2026hif} and the combined COHERENT and reactor experiments ($\chi^2_{\text{SM}} + 6$) \cite{Rathsman:2026smv} in the ($C_\text{eff}^{\nu_e}$, $C_\text{eff}^{\nu_\mu}$) parameter space, as shown in Fig.~\ref{fig:limit_2dof_comp}. These prior studies explore a similar effective coupling framework within the context of CE$\nu$NS, offering highly relevant benchmarks for our findings. However, we stress a critical methodological distinction: while our analysis robustly marginalizes over the remaining flavor coupling ($C_\text{eff}^{\nu_\tau}$) to account for all flavor correlations, the referenced limits were derived without such full marginalization.

To ensure a statistically consistent comparison across different experimental methodologies, we plot our global analysis contours at both 90\% CL (red hatched region) and the $\chi^2_{\text{SM}} + 6$ threshold (blue hatched region). As visually evident, our combined analysis of the XENONnT, PandaX-4T, and LUX-ZEPLIN datasets provides remarkably competitive constraints. Our global contours drastically restrict the extensive, unconstrained diagonal swath left by IceCube and significantly carve into the broader allowed region of the reactor and COHERENT datasets. 
This substantial reduction in the viable parameter space highlights the immense and previously untapped potential of dark matter direct detection experiments. By leveraging the CE$\nu$NS signals induced by $^8$B solar neutrinos, these multi-ton facilities offer a powerful and complementary new avenue for probing the hypothetical $X_{17}$ mediator and rigorously testing its viability as an explanation for the Atomki anomaly.

\section{Conclusions}\label{sec:conc} 
Motivated by the first observation of coherent elastic neutrino-nucleus scattering induced by $^8$B solar neutrinos, we have derived robust constraints on the effective couplings of a light $Z'$ gauge boson, interpreted here as the $X_{17}$ particle. Initially conjectured to explain the invariant mass anomalies observed by the Atomki experiment, this hypothetical mediator can be handled within a $U(1)'$ gauge symmetry framework. Crucially, we have explicitly accounted for the flavor-dependent nature of the couplings, which naturally arise from solar neutrino flavor transitions during their propagation to Earth.

Our analysis utilized the latest datasets from direct detection searches at XENONnT, PandaX-4T, and LUX-ZEPLIN experiments. While these multi-ton direct detection facilities were fundamentally designed to search for weakly interacting massive dark matter particles, their recent milestone in detecting solar neutrino-induced nuclear recoils offers a novel and powerful avenue for probing BSM physics. We systematically evaluated the flavor-dependent effective couplings—$C_\text{eff}^{\nu_e}$, $C_\text{eff}^{\nu_\mu}$, and $C_\text{eff}^{\nu_\tau}$—that quantify the $X_{17}$ interactions with the xenon target. By rigorously marginalizing over the relevant effective neutrino couplings, we have derived comprehensive 1~dof and 2~dof limits of the effective vector couplings using each dataset released by the experiments. Our global analysis reveals that PandaX-4T and LUX-ZEPLIN provide relatively similar behavior, while the combined datasets effectively break individual experimental degeneracies. Most importantly, we have found that the SM hypothesis remains completely protected and consistent within the 90\% CL allowed regions across all flavor combinations.
Overall, our derived bounds are highly competitive with the allowed parameter spaces previously mapped by IceCube and COHERENT+reactor studies. We also highlight the sensitivity of these detectors to the tau-flavor coupling, a direct consequence of the oscillated solar neutrino flux. 

These findings conclusively demonstrate that direct detection experiments play a compelling and complementary role in the ongoing search for the $X_{17}$ particle. Therefore, currently operating multi-ton liquid xenon facilities serve as formidable testing grounds for the $X_{17}$ hypothesis. Our comprehensive limits could provide hints for future theoretical and experimental endeavors.
%important theoretical and experimental hints for future endeavors.

\section*{Acknowledgments}
%The authors would like to thank reviewers for their valuable comments and fruitful discussions. 
This work was supported by the Scientific and Technological Research Council of Türkiye (TUBITAK) under the project numbers 124F416 and 126F009. We would like to acknowledge the networking support of the COST Action CA21106 - COSMIC WISPers in the Dark Universe: Theory, astrophysics and experiments (CosmicWISPers).
%\newpage
\bibliographystyle{unsrt}

\begin{thebibliography}{99}
	\bibitem{Gulyas:2015mia}
	J.~Guly{\'a}s, T.~J.~Ketel, A.~J.~Krasznahorkay, M.~Csatl{\'o}s, L.~Csige, Z.~G{\'a}csi, M.~Hunyadi, A.~Krasznahorkay, A.~Vit{\'e}z and T.~G.~Tornyi,
	%``A pair spectrometer for measuring multipolarities of energetic nuclear transitions,''
	\href{https://doi.org/10.1016/j.nima.2015.11.009}{Nucl. Instrum. Meth. A \textbf{808} (2016), 21}.
	%[arXiv:1504.00489 [nucl-ex]].
	
	\bibitem{Krasznahorkay:2015iga}
	A.~J.~Krasznahorkay, M.~Csatl{\'o}s, L.~Csige, Z.~G{\'a}csi, J.~Guly{\'a}s, M.~Hunyadi, T.~J.~Ketel, A.~Krasznahorkay, I.~Kuti and B.~M.~Nyak{\'o}, \textit{et al.}
	%``Observation of Anomalous Internal Pair Creation in Be8 : A Possible Indication of a Light, Neutral Boson,''
	\href{https://doi.org/10.1103/PhysRevLett.116.042501}{Phys. Rev. Lett. \textbf{116} (2016), 042501}.
	%[arXiv:1504.01527 [nucl-ex]].
	
	\bibitem{Krasznahorkay:2021joi}
	A.~J.~Krasznahorkay, M.~Csatl{\'o}s, L.~Csige, J.~Guly{\'a}s, A.~Krasznahorkay, B.~M.~Nyak{\'o}, I.~Rajta, J.~Tim{\'a}r, I.~Vajda and N.~J.~Sas,
	%``New anomaly observed in He4 supports the existence of the hypothetical X17 particle,''
	\href{https://doi.org/10.1103/PhysRevC.104.044003}{Phys. Rev. C \textbf{104} (2021), 044003}.
	%[arXiv:2104.10075 [nucl-ex]].
	
	\bibitem{Krasznahorkay:2022pxs}
	A.~J.~Krasznahorkay, A.~Krasznahorkay, M.~Begala, M.~Csatl{\'o}s, L.~Csige, J.~Guly{\'a}s, A.~Krak{\'o}, J.~Tim{\'a}r, I.~Rajta and I.~Vajda, \textit{et al.}
	%``New anomaly observed in C12 supports the existence and the vector character of the hypothetical X17 boson,''
	\href{https://doi.org/10.1103/PhysRevC.106.L061601}{Phys. Rev. C \textbf{106} (2022), L061601}.
	%[arXiv:2209.10795 [nucl-ex]].
	
	\bibitem{Anh:2024req}
	T.~Anh, T.~D.~Trong, A.~J.~Krasznahorkay, A.~Krasznahorkay, J.~Moln{\'a}r, Z.~Pintye, N.~A.~Viet, N.~Nghia, D.~T.~K.~Linh and B.~T.~Hoa, \textit{et al.}
	%``Checking the $^{8}$Be Anomaly with a Two-Arm Electron Positron Pair Spectrometer,''
	\href{https://doi.org/10.3390/universe10040168}{Universe \textbf{10} (2024), 168}.
	%[arXiv:2401.11676 [nucl-ex]].
	
	\bibitem{Abraamyan:2024bbs}
	K.~U.~Abraamyan, C.~Austin, M.~I.~Baznat, K.~K.~Gudima, M.~A.~Kozhin, S.~G.~Reznikov and A.~S.~Sorin,
	%``Observation of Structures at {\textasciitilde}17 and {\textasciitilde}38 MeV/c$^{2}$ in the {\ensuremath{\gamma}}{\ensuremath{\gamma}} Invariant Mass Spectrum in dCu Collisions at a Momentum of 3.8 GeV/c per Nucleon,''
	\href{https://doi.org/10.1134/S1063779624700412}{Phys. Part. Nucl. \textbf{55} (2024), 868}.
	
	\bibitem{PADME:2025dla}
	F.~Bossi \textit{et al.} (PADME Collaboration),
	%``Search for a new 17 MeV resonance via e$^{+}$e$^{−}$ annihilation with the PADME experiment,''
	\href{https://doi.org/10.1007/JHEP11(2025)007}{JHEP \textbf{11} (2025), 007}.
	%[arXiv:2505.24797 [hep-ex]].

    \bibitem{MEGII:2024urz}
	K.~Afanaciev \textit{et al.} (MEG II Collaboration),
	%``Search for the X17 particle in $^{7}\textrm{Li}(\textrm{p},\textrm{e}^+ \textrm{e}^{-}) ^{8}\textrm{Be}$ processes with the MEG II detector,''
	\href{https://doi.org/10.1140/epjc/s10052-025-14345-0}{Eur. Phys. J. C \textbf{85} (2025), 763}.
	%[arXiv:2411.07994 [nucl-ex]].
    
	\bibitem{Zhang:2017zap}
	X.~Zhang and G.~A.~Miller,
	%``Can nuclear physics explain the anomaly observed in the internal pair production in the Beryllium-8 nucleus?,''
	\href{https://doi.org/10.1016/j.physletb.2017.08.013}{Phys. Lett. B \textbf{773} (2017), 159-165}.
	%[arXiv:1703.04588 [nucl-th]].
	
	\bibitem{Koch:2020ouk}
	B.~Koch,
	%``X17: A new force, or evidence for a hard $\gamma +\gamma$ process?,''
	\href{https://doi.org/10.1016/j.nuclphysa.2021.122143}{Nucl. Phys. A \textbf{1008} (2021), 122143}.
	%[arXiv:2003.05722 [hep-ph]].
	
	\bibitem{Hayes:2021hin}
	A.~C.~Hayes, J.~L.~Friar, G.~M.~Hale and G.~T.~Garvey,
	%``Angular correlations in the e+e{\ensuremath{-}} decay of excited states in Be8,''
	\href{https://doi.org/10.1103/PhysRevC.105.055502}{Phys. Rev. C \textbf{105} (2022), 055502}.
	%[arXiv:2106.06834 [nucl-th]].
	
	\bibitem{Viviani:2024czq}
	M.~Viviani, E.~Filandri, L.~Girlanda, C.~Gustavino, A.~Kievsky and L.~E.~Marcucci,
	%``X17 boson and the H2(p,e+e{\ensuremath{-}})He3 and H2(n,e+e{\ensuremath{-}})H3 processes: A theoretical analysis,''
	\href{https://doi.org/10.1103/PhysRevC.111.034002}{Phys. Rev. C \textbf{111} (2025), 034002}.
	%[arXiv:2408.16744 [nucl-th]].
	
	\bibitem{Feng:2016jff}
	J.~L.~Feng, B.~Fornal, I.~Galon, S.~Gardner, J.~Smolinsky, T.~M.~P.~Tait and P.~Tanedo,
	%``Protophobic Fifth-Force Interpretation of the Observed Anomaly in $^8$Be Nuclear Transitions,''
	\href{https://doi.org/10.1103/PhysRevLett.117.071803}{Phys. Rev. Lett. \textbf{117} (2016), 071803}.
	%[arXiv:1604.07411 [hep-ph]].
	
	\bibitem{Feng:2016ysn}
	J.~L.~Feng, B.~Fornal, I.~Galon, S.~Gardner, J.~Smolinsky, T.~M.~P.~Tait and P.~Tanedo,
	%``Particle physics models for the 17 MeV anomaly in beryllium nuclear decays,''
	\href{https://doi.org/10.1103/PhysRevD.95.035017}{Phys. Rev. D \textbf{95} (2017), 035017}.
	%[arXiv:1608.03591 [hep-ph]].
	
	\bibitem{Feng:2020mbt}
	J.~L.~Feng, T.~M.~P.~Tait and C.~B.~Verhaaren,
	%``Dynamical Evidence For a Fifth Force Explanation of the ATOMKI Nuclear Anomalies,''
	\href{https://doi.org/10.1103/PhysRevD.102.036016}{Phys. Rev. D \textbf{102} (2020), 036016}.
	%[arXiv:2006.01151 [hep-ph]].
	
	
    \bibitem{Kozaczuk:2016nma}
    J.~Kozaczuk, D.~E.~Morrissey and S.~R.~Stroberg,
    %``Light axial vector bosons, nuclear transitions, and the $^8$Be anomaly,''
    \href{https://doi.org/10.1103/PhysRevD.95.115024}{Phys. Rev. D \textbf{95} (2017), 115024}.
    %[arXiv:1612.01525 [hep-ph]].
    
    \bibitem{Barducci:2022lqd}
    D.~Barducci and C.~Toni,
    %``An updated view on the ATOMKI nuclear anomalies,''
    \href{https://doi.org/10.1007/JHEP02(2023)154}{JHEP \textbf{02} (2023), 154 [erratum: JHEP \textbf{07} (2023), 168]}.
    %[arXiv:2212.06453 [hep-ph]].

    \bibitem{Hostert:2023tkg}
    M.~Hostert and M.~Pospelov,
    %``Pion decay constraints on exotic 17~MeV vector bosons,''
    \href{https://doi.org/10.1103/PhysRevD.108.055011}{Phys. Rev. D \textbf{108} (2023), 055011}.
    %[arXiv:2306.15077 [hep-ph]].
    
    \bibitem{Mommers:2024qzy}
    C.~J.~G.~Mommers and M.~Vanderhaeghen,
    %``Constraining the axial-vector X17 interpretation with 12C data,''
    \href{https://doi.org/10.1016/j.physletb.2024.139031}{Phys. Lett. B \textbf{858} (2024), 139031}.
    %[arXiv:2406.08143 [hep-ph]].
    
    
    \bibitem{Fieg:2026zkg}
    M.~H.~Fieg, T.~M{\"a}kel{\"a}, T.~M.~P.~Tait and M.~Toman,
    %``The X17 with Chiral Couplings,''
    \href{https://arxiv.org/abs/2602.11263}{[arXiv:2602.11263 [hep-ph]]}.
    
    \bibitem{Jiang:2026snq}
    J.~Jiang, C.~F.~Qiao and Y.~H.~Zhao,
    %``Where to find $X(17)$?,''
    \href{https://arxiv.org/abs/2601.08567}{[arXiv:2601.08567 [hep-ph]]}.

    
    
    \bibitem{DelleRose:2017xil}
    L.~Delle Rose, S.~Khalil and S.~Moretti,
    %``Explanation of the 17 MeV Atomki anomaly in a U(1)' -extended two Higgs doublet model,''
    \href{https://doi.org/10.1103/PhysRevD.96.115024}{Phys. Rev. D \textbf{96} (2017), 115024}.
    %[arXiv:1704.03436 [hep-ph]].
    
    \bibitem{DelleRose:2018eic}
    L.~Delle Rose, S.~Khalil, S.~J.~D.~King, S.~Moretti and A.~M.~Thabt,
    %``Atomki Anomaly in Family-Dependent $U(1)'$ Extension of the Standard Model,''
    \href{https://doi.org/10.1103/PhysRevD.99.055022}{Phys. Rev. D \textbf{99} (2019), 055022}.
    %[arXiv:1811.07953 [hep-ph]].
        
    \bibitem{Pulice:2019xel}
    B.~Puli{\c{c}}e,
    %``A Family-nonuniversal U(1)' Model for Excited Beryllium Decays,''
    \href{https://doi.org/10.1016/j.cjph.2020.11.021}{Chin. J. Phys. \textbf{71} (2021), 506}.
    %[arXiv:1911.10482 [hep-ph]].

    \bibitem{Cerdeno:2016sfi}
    D.~G.~Cerde{\~n}o, M.~Fairbairn, T.~Jubb, P.~A.~N.~Machado, A.~C.~Vincent and C.~B{\oe}hm,
    %``Physics from solar neutrinos in dark matter direct detection experiments,''
    \href{https://doi.org/10.1007/JHEP09(2016)048}{JHEP \textbf{05} (2016), 118 [erratum: JHEP \textbf{09} (2016), 048]}.
    %[arXiv:1604.01025 [hep-ph]].

    \bibitem{Boehm:2020ltd}
    C.~Boehm, D.~G.~Cerdeno, M.~Fairbairn, P.~A.~N.~Machado and A.~C.~Vincent,
    %``Light new physics in XENON1T,''
    \href{https://doi.org/10.1103/PhysRevD.102.115013}{Phys. Rev. D \textbf{102} (2020), 115013}.
    %[arXiv:2006.11250 [hep-ph]].


    \bibitem{Demirci:2026nju}
    M.~Demirci and M.~F.~Mustamin,
    %``Solar neutrino probes of light new physics: Updated limits from LUX-ZEPLIN experiment,''
    \href{https://doi.org/10.1103/3cn3-rp6j}{Phys. Rev. D \textbf{113} (2026), 055036}.
    %[arXiv:2603.19467 [hep-ph]].
        
    \bibitem{Freedman:1973yd}
    D.~Z.~Freedman,
    %``Coherent Neutrino Nucleus Scattering as a Probe of the Weak Neutral Current,''
    \href{https://doi.org/10.1103/PhysRevD.9.1389}{Phys. Rev. D \textbf{9} (1974), 1389}.
    
    \bibitem{Drukier:1984vhf}
    A.~Drukier and L.~Stodolsky,
    %``Principles and Applications of a Neutral Current Detector for Neutrino Physics and Astronomy,''
    \href{https://doi.org/10.1103/PhysRevD.30.2295}{Phys. Rev. D \textbf{30} (1984), 2295}.
    
    \bibitem{COHERENT:2017ipa}
    D.~Akimov \textit{et al.} (COHERENT Collaboration),
    %``Observation of Coherent Elastic Neutrino-Nucleus Scattering,''
    \href{https://doi.org/10.1126/science.aao0990}{Science \textbf{357} (2017), 1123}.
    %[arXiv:1708.01294 [nucl-ex]].
    
    \bibitem{COHERENT:2020iec}
    D.~Akimov \textit{et al.} (COHERENT Collaboration),
    %``First Measurement of Coherent Elastic Neutrino-Nucleus Scattering on Argon,''
    \href{https://doi.org/10.1103/PhysRevLett.126.012002}{Phys. Rev. Lett. \textbf{126} (2021), 012002}.
    %[arXiv:2003.10630 [nucl-ex]].
    
    \bibitem{COHERENT:2024axu}
    S.~Adamski \textit{et al.} (COHERENT Collaboration),
    %``Evidence of Coherent Elastic Neutrino-Nucleus Scattering with COHERENT{\textquoteright}s Germanium Array,''
    \href{https://doi.org/10.1103/PhysRevLett.134.231801}{Phys. Rev. Lett. \textbf{134} (2025), 231801}.
    %[arXiv:2406.13806 [hep-ex]].
    
    \bibitem{Colaresi:2022obx}
    J.~Colaresi, J.~I.~Collar, T.~W.~Hossbach, C.~M.~Lewis and K.~M.~Yocum,
    %``Measurement of Coherent Elastic Neutrino-Nucleus Scattering from Reactor Antineutrinos,''
    \href{https://doi.org/10.1103/PhysRevLett.129.211802}{Phys. Rev. Lett. \textbf{129} (2022), 211802}.
    %[arXiv:2202.09672 [hep-ex]].
    
    \bibitem{Ackermann:2025obx}
    N.~Ackermann, H.~Bonet, A.~Bonhomme, C.~Buck, K.~F{\"u}lber, J.~Hakenm{\"u}ller, J.~Hempfling, G.~Heusser, M.~Lindner and W.~Maneschg, \textit{et al.}
    %``Direct observation of coherent elastic antineutrino{\textendash}nucleus scattering,''
    \href{https://doi.org/10.1038/s41586-025-09322-2}{Nature \textbf{643} (2025), 1229}.
	%[arXiv:2501.05206 [hep-ex]].
    
    \bibitem{Denton:2018xmq}
    P.~B.~Denton, Y.~Farzan and I.~M.~Shoemaker,
    %``Testing large non-standard neutrino interactions with arbitrary mediator mass after COHERENT data,''
    \href{https://doi.org/10.1007/JHEP07(2018)037}{JHEP \textbf{07} (2018), 037}.
    %[arXiv:1804.03660 [hep-ph]].
    
    \bibitem{Khan:2019cvi}
    A.~N.~Khan and W.~Rodejohann,
    %``New physics from COHERENT data with an improved quenching factor,''
    \href{https://doi.org/10.1103/PhysRevD.100.113003}{Phys. Rev. D \textbf{100} (2019), 113003}.
    %[arXiv:1907.12444 [hep-ph]].
    
    \bibitem{AristizabalSierra:2022axl}
    D.~Aristizabal Sierra, V.~De Romeri and D.~K.~Papoulias,
    %``Consequences of the Dresden-II reactor data for the weak mixing angle and new physics,''
    \href{https://doi.org/10.1007/JHEP09(2022)076}{JHEP \textbf{09} (2022), 076}.
    %[arXiv:2203.02414 [hep-ph]].
    
    \bibitem{AtzoriCorona:2022moj}
    M.~Atzori Corona, M.~Cadeddu, N.~Cargioli, F.~Dordei, C.~Giunti, Y.~F.~Li, E.~Picciau, C.~A.~Ternes and Y.~Y.~Zhang,
    %``Probing light mediators and (g {\ensuremath{-}} 2)$_{μ}$ through detection of coherent elastic neutrino nucleus scattering at COHERENT,''
    \href{https://doi.org/10.1007/JHEP05(2022)109}{JHEP \textbf{05} (2022), 109}.
    %[arXiv:2202.11002 [hep-ph]].
    
    \bibitem{Demirci:2023tui}
    M.~Demirci and M.~F.~Mustamin,
    %``Solar neutrino constraints on light mediators through coherent elastic neutrino-nucleus scattering,''
    \href{https://doi.org/10.1103/PhysRevD.109.015021}{Phys. Rev. D \textbf{109} (2024), 015021}.
    %[arXiv:2312.17502 [hep-ph]].
    
    \bibitem{DeRomeri:2025nkx}
    V.~De Romeri, A.~Majumdar, D.~K.~Papoulias and R.~Srivastava,
    %``New light mediators and the neutrino fog: Implications from XENONnT nuclear recoil data,''
    \href{https://doi.org/10.1088/1475-7516/2026/05/093}{JCAP \textbf{05} (2026), 093}.
    %[arXiv:2512.08853 [hep-ph]].
    
    \bibitem{Denton:2023gat}
    P.~B.~Denton and J.~Gehrlein,
    %``Neutrino constraints and the ATOMKI X17 anomaly,''
    \href{https://doi.org/10.1103/PhysRevD.108.015009}{Phys. Rev. D \textbf{108} (2023), 015009}.
    %[arXiv:2304.09877 [hep-ph]].
    
    \bibitem{Cederkall:2025bka}
    J.~Cederk{\"a}ll, Y.~Hi{\c{c}}y{\i}lmaz, E.~Lytken, S.~Moretti and J.~Rathsman,
    %``Hunting the elusive X17 in CE{\ensuremath{\nu}}NS at the ESS,''
    \href{https://doi.org/10.1007/JHEP12(2025)027}{JHEP \textbf{12} (2025), 027}.
    %[arXiv:2509.15128 [hep-ph]].
    
    \bibitem{Garoby:2017vew}
    R.~Garoby, H.~Danared, I.~Alonso, E.~Bargallo, B.~Cheymol, C.~Darve, M.~Eshraqi, H.~Hassanzadegan, A.~Jansson and I.~Kittelmann, \textit{et al.}
    %``The European Spallation Source Design,''
    \href{https://doi.org/10.1088/1402-4896/aa9bff}{Phys. Scripta \textbf{93} (2018), 014001}.
    
    \bibitem{Baxter:2019mcx}
    D.~Baxter, J.~I.~Collar, P.~Coloma, C.~E.~Dahl, I.~Esteban, P.~Ferrario, J.~J.~Gomez-Cadenas, M.~C.~Gonzalez-Garcia, A.~R.~L.~Kavner and C.~M.~Lewis, \textit{et al.}
    %``Coherent Elastic Neutrino-Nucleus Scattering at the European Spallation Source,''
    \href{https://doi.org/10.1007/JHEP02(2020)123}{JHEP \textbf{02} (2020), 123}.
    %[arXiv:1911.00762 [physics.ins-det]].
    
    \bibitem{Abele:2022iml}
    H.~Abele, A.~Alekou, A.~Algora, K.~Andersen, S.~Bae{\ss}ler, L.~Barron-P{\'a}los, J.~Barrow, E.~Baussan, P.~Bentley and Z.~Berezhiani, \textit{et al.}
    %``Particle Physics at the European Spallation Source,''
    \href{https://doi.org/10.1016/j.physrep.2023.06.001}{Phys. Rept. \textbf{1023} (2023), 1}.
    %[arXiv:2211.10396 [physics.ins-det]].
    
    \bibitem{XENON:2024ijk}
    E.~Aprile \textit{et al.} (XENON Collaboration),
    %``First Indication of Solar B8 Neutrinos via Coherent Elastic Neutrino-Nucleus Scattering with XENONnT,''
    \href{https://doi.org/10.1103/PhysRevLett.133.191002}{Phys. Rev. Lett. \textbf{133} (2024), 191002}.
    %[arXiv:2408.02877 [hep-ex]].
    
    \bibitem{PandaX:2024muv}
    Z.~Bo \textit{et al.} (PandaX Collaboration),
    %``First Indication of Solar B8 Neutrinos through Coherent Elastic Neutrino-Nucleus Scattering in PandaX-4T,''  
    \href{https://doi.org/10.1103/PhysRevLett.133.191001}{Phys. Rev. Lett. \textbf{133} (2024), 191001}.
    %[arXiv:2407.10892 [hep-ex]].
    
    \bibitem{LZ:2025igz}
    D.~S.~Akerib \textit{et al.} (LZ Collaboration),
    %``Searches for Light Dark Matter and Evidence of Coherent Elastic Neutrino-Nucleus Scattering of Solar Neutrinos with the LUX-ZEPLIN (LZ) Experiment,''
    \href{https://lss.fnal.gov/archive/2025/pub/fermilab-pub-25-0917-v.pdf}{arXiv:2512.08065 [hep-ex]}.
       
    
    \bibitem{Erler:2013xha}
    J.~Erler and S.~Su,
    %``The Weak Neutral Current,''
    \href{https://doi.org/10.1016/j.ppnp.2013.03.004}{Prog. Part. Nucl. Phys. \textbf{71} (2013), 119-149}.
    %[arXiv:1303.5522 [hep-ph]].
    
    \bibitem{AtzoriCorona:2025xwr}
    M.~Atzori Corona, M.~Cadeddu, N.~Cargioli, F.~Dordei, C.~Giunti and C.~A.~Ternes,
    %``Standard Model Tested with Neutrinos,''
    \href{https://doi.org/10.1103/dplq-dvc8}{Phys. Rev. Lett. \textbf{135} (2025), 231803}.
    %[arXiv:2504.05272 [hep-ph]].
             
    
    \bibitem{Klein:1999qj}
    S.~Klein and J.~Nystrand,
    %``Exclusive vector meson production in relativistic heavy ion collisions,''
    \href{https://doi.org/10.1103/PhysRevC.60.014903}{Phys. Rev. C \textbf{60} (1999), 014903}. 
    %[arXiv:hep-ph/9902259 [hep-ph]].

    \bibitem{Berglund:2009}
    Berglund, Michael, and Wieser, Michael E.,
    %"Isotopic compositions of the elements 2009 (IUPAC Technical Report)," 
    \href{https://doi.org/10.1351/PAC-REP-10-06-02}{Pure and Applied Chemistry, 83 (2011), 397.}   
    
    
    \bibitem{Bahcall:1989ks}
    J.~N.~Bahcall,
    %``NEUTRINO ASTROPHYSICS,''
    \href{https://www.cambridge.org/tr/universitypress/subjects/physics/astrophysics/neutrino-astrophysics}{Cambridge University Press (1989).} %ISBN: 9780521379755}.
    
    \bibitem{Vinyoles:2016djt}
    N.~Vinyoles, A.~M.~Serenelli, F.~L.~Villante, S.~Basu, J.~Bergstr{\"o}m, M.~C.~Gonzalez-Garcia, M.~Maltoni, C.~Pe{\~n}a-Garay and N.~Song,
    %``A new Generation of Standard Solar Models,''
    \href{https://doi.org/10.3847/1538-4357/835/2/202}{Astrophys. J. \textbf{835} (2017), 202}.
    %[arXiv:1611.09867 [astro-ph.SR]].
    
    \bibitem{Maltoni:2015kca}
    M.~Maltoni and A.~Y.~Smirnov,
    %``Solar neutrinos and neutrino physics,''
    \href{https://doi.org/10.1140/epja/i2016-16087-0}{Eur. Phys. J. A \textbf{52} (2016), 87}.
    %[arXiv:1507.05287 [hep-ph]].
    
    \bibitem{Esteban:2024eli}
    I.~Esteban, M.~C.~Gonzalez-Garcia, M.~Maltoni, I.~Martinez-Soler, J.~P.~Pinheiro and T.~Schwetz,
    %``NuFit-6.0: updated global analysis of three-flavor neutrino oscillations,''
    \href{https://doi.org/10.1007/JHEP12(2024)216}{JHEP \textbf{12} (2024), 216}.
    %[arXiv:2410.05380 [hep-ph]].
    
    \bibitem{Sarkis:2026lds}
    Y.~Sarkis, J.~C.~D'Olivo and A.~Aguilar-Arevalo,
    %``Light and charge yield in noble liquids from sub-keV to MeV: a first-principles approach beyond Lindhard,''
    \href{https://doi.org/10.1088/1748-0221/21/04/C04036}{JINST \textbf{21} (2026), C04036}.
    
    \bibitem{XENON:2024xgd}
    E.~Aprile \textit{et al.} (XENON Collaboration){\textdaggerdbl}{\textdaggerdbl}],
    %``XENONnT WIMP search: Signal and background modeling and statistical inference,''
    \href{https://doi.org/10.1103/PhysRevD.111.103040}{Phys. Rev. D \textbf{111} (2025), 103040}.
    %[arXiv:2406.13638 [physics.data-an]].
    
    \bibitem{Baker:1983tu}
    S.~Baker and R.~D.~Cousins,
    %``Clarification of the Use of Chi Square and Likelihood Functions in Fits to Histograms,''
    \href{https://doi.org/10.1016/0167-5087(84)90016-4}{Nucl. Instrum. Meth. \textbf{221} (1984), 437}.
    
    \bibitem{Fogli:2002pt}
    G.~L.~Fogli, E.~Lisi, A.~Marrone, D.~Montanino and A.~Palazzo,
    %``Getting the most from the statistical analysis of solar neutrino oscillations,''   
    \href{https://doi.org/10.1103/PhysRevD.66.053010}{ Phys. Rev. D \textbf{66} (2002), 053010}.
    %[arXiv:hep-ph/0206162 [hep-ph]].
    
    \bibitem{Baxter:2021pqo}
    D.~Baxter, I.~M.~Bloch, E.~Bodnia, X.~Chen, J.~Conrad, P.~Di Gangi, J.~E.~Y.~Dobson, D.~Durnford, S.~J.~Haselschwardt and A.~Kaboth, \textit{et al.}
    %``Recommended conventions for reporting results from direct dark matter searches,''
    \href{https://doi.org/10.1140/epjc/s10052-021-09655-y}{Eur. Phys. J. C \textbf{81} (2021), 907}.
    %[arXiv:2105.00599 [hep-ex]].

    \bibitem{Rathsman:2026smv}
    J.~Rathsman, J.~Cederk{\"a}ll, Y.~Hicyilmaz, E.~Lytken and S.~Moretti,
    %``Glimpses of the X17 from coherent elastic neutrino nucleus scattering,''
    \href{https://arxiv.org/abs/2603.15246}{[arXiv:2603.15246 [hep-ph]]}.
    
    \bibitem{Rathsman:2026hif}
    J.~Rathsman, J.~Cederk{\"a}ll, Y.~Hicyilmaz, E.~Lytken and S.~Moretti,
    %``The X17 Existence Hinted at by Nuclear Reactor Neutrinos,''
    \href{https://arxiv.org/abs/2605.10689}{[arXiv:2605.10689 [hep-ph]]}.

    
\end{thebibliography}

\end{document}